\documentclass[intlimits,twoside,a4paper]{article}

\usepackage{amsmath,amssymb}
\usepackage{graphicx}
\usepackage{wrapfig}

\usepackage[T2A]{fontenc}
\usepackage[cp1251]{inputenc}
\usepackage{cmpj}

\newcommand{\eps}{\varepsilon}

\issue{2011}{14}{4}{43602}

\doinumber{10.5488/CMP.14.43602}

\title[Mitsui model with diagonal strains]{Mitsui model with diagonal strains: A unified description of external pressure effect and thermal expansion
of Rochelle salt NaKC$_4$H$_4$O$_6\cdot4$H$_2$O }

\author[A.P. Moina, R.R. Levitskii,
I.R. Zachek]{A.P. Moina \refaddr{label1}, R.R. Levitskii
\refaddr{label1}, I.R. Zachek \refaddr{label2}}

\authorcopyright{A.P. Moina, R.R. Levitskii,
I.R. Zachek, 2011}
\addresses{
\addr{label1} Institute for Condensed Matter Physics
of the National Academy of Sciences of Ukraine,\\
1 Svientsitskii Str., 79011 Lviv, Ukraine
 \addr{label2} Lviv Polytechnical
National University, 12 Bandera Str., 79013 Lviv, Ukraine}

\date{Received August 29, 2011, in final form November 9, 2011}
\begin{document}

\maketitle

\begin{abstract}

We elaborate a  modification of the deformable two-sublattice
Mitsui model of [Levitskii~R.R. et al., Phys. Rev.~B. 2003,
\textbf{67}, 174112] and [Levitskii~R.R. et al.,  Condens. Matter
Phys., 2005, \textbf{8}, 881] that consistently takes into account
diagonal components of the strain tensor, arising either due to
external pressures or due to thermal expansion. We calculate the
related to those strains thermal, piezoelectric, and elastic
characteristics of the system. Using the developed fitting
procedure, a set of the model parameters is found for the case
of Rochelle salt crystals, providing a satisfactory agreement with
the available experimental data for the hydrostatic and uniaxial
pressure dependences of the Curie temperatures, temperature
dependences of spontaneous diagonal strains, linear thermal
expansion coefficients, elastic constants $c_{ij}^E $ and
$c_{i4}^E $, piezoelectric coefficients $d_{1i}$ and $g_{1i}$
($i=1,2,3$). The hydrostatic pressure variation of dielectric
permittivity is described using a derived expression for the
permittivity of a partially clamped crystal. The dipole moments
and the asymmetry parameter of Rochelle salt are found to increase
with hydrostatic pressure.

\keywords Rochelle salt, thermal expansion, hydrostatic pressure,
uniaxial pressure, Mitsui model
 \pacs 65.40.De, 77.80.B-, 77.65.Bn, 65.40.Ba, 77.22.Ch
\end{abstract}

\maketitle

\section{Introduction}
The Mitsui model \cite{80} (two-sublattice Ising model with
asymmetric potentials) has been originally proposed for
description of the reentrant phase behavior in Rochelle salt
crystals. It considers the motion of certain ordering units in two
interpenetrating sublattices of asymmetric double-well potentials.
The model with certain modifications is also applicable to several
other ferroelectrics \cite{vaks-review}, like those of the
Rochelle salt type (deuterated and ammonium-doped \cite{Lunk}
Rochelle salt at least at low doping ), RbHSO$_4$ type
\cite{Alexandrov,Blat}, AgNa(NO$_2)_2$ (SSN
\cite{watarai1,watarai2}), SASD type \cite{sasd}
(NaNH$_4$SO$_4$$\cdot$2H$_2$O and NaNH$_4$SeO$_4$$\cdot$2H$_2$O),
etc.

Very often, inclusion of deformational effects into the Mitsui
model is indispensable for a proper description of the system
behavior even at ambient pressure. Thus, a strong piezoelectricity
associated with polarization $P_1$ and shear strain $\eps_4$
essentially affects the dynamic dielectric response of Rochelle
salt due to the effect of crystal clamping by the high-frequency
measuring field. The conventional Mitsui model yields a
qualitatively incorrect behavior of the relaxation time and
dynamic dielectric permittivity near the Curie temperatures. This
problem is resolved \cite{ourrs} by taking into account the
piezoelectric coupling with $\eps_4$, also permitting to describe
the phenomena of piezoelectric resonance and sound attenuation
\cite{ourrs2}.

The sequence of phase transitions observed in NH$_4$HSO$_4$ crystals can be described within the mean field approximation
for the Mitsui model only with  temperature dependent interaction
constants \cite{Blat}. It means that one must take into account
the effect of thermal expansion, which is, ultimately, a
deformational effect.

Since high pressure studies are the only means to continuously vary
the system geometrical parameters, such as the orientation angles
of atomic groups that form the dipole moments, interatomic
distances, hydrogen bonds parameters, etc, as well as the
interparticle interactions, and other parameters of the system,
they can provide a valuable information on the mechanism of the
phase transitions in ferroelectric crystals.  A better insight is
obtained if the effects of hydrostatic and various uniaxial and
biaxial pressures are explored. Irrespectively of the crystal
symmetry, these pressures produce diagonal components of the
lattice strain tensor $\eps_i$ ($i=1,2,3$). In low-symmetry
systems, shear strains $\eps_j$ ($j=4,5,6$) can be induced as
well. The diagonal strains also arise due to the thermal expansion
of the crystals.

Nowadays, Rochelle salt
\cite{YadlovkerBerger,YadlovkerBerger2,Baryshnikov1,Baryshnikov2}
and other crystals  \cite{rbhso4nano} described by the Mitsui
model often serve as test materials for experiments with nanosize
phenomena. Properties of the ferroelectric nano-inclusions in a
solid matrix are strongly affected by surface tension and thermal
mismatch stresses. Thus, radial stresses in the plane
perpendicular to spontaneous polarization arise in Rochelle salt
nanorods grown in pores of alumina films
\cite{YadlovkerBerger,YadlovkerBerger2}, inducing diagonal strains
only. A phenomenological theory of ferroelectric properties of
such nanorods was presented in \cite{Morozovska}.

The simplest way to incorporate an external pressure into a spin
model is to consider its parameters (interaction constants)
phenomenologically as linear functions of pressure. This approach
has been used for different versions of the Mitsui model to
describe the hydrostatic pressure variation of the transition
temperatures in Rochelle salt  \cite{AAY} and  SASD
\cite{SASD-pressure}. However, to describe a uniaxial stress
dependence of $T_{\rm C}$ in the same way, one would have to find
new values of the fitting parameters for each stress direction. A
unified model description of hydrostatic and uniaxial pressure
effects, and, at the same time, the crystal thermal expansion at
ambient pressure, with a single set of the fitting parameters is
possible if we include into the model the lattice diagonal
strains, instead of the pressures. Such a microscopic-like model
of bulk crystals that includes the diagonal strains will be very
helpful when one develops a model description of the above mentioned
 nanocrystal behavior.

The goal of the present paper is to develop a modification of the
deformable Mitsui model with the shear strain $\eps_4$
\cite{ourrs}, which would also take into account the diagonal
components of the lattice strain tensor. The first attempt to
create such a modification was made in \cite{monoclinic}. The
interaction constants and the asymmetry parameter were taken to be
linear functions of the diagonal strains. Expressions for the
piezoelectric and elastic characteristics of Rochelle salt,
associated with these strains, have been obtained. However, all
actual calculations were performed in the approximation of zero
thermal strains; the external hydrostatic or uniaxial pressure
effect was not considered, and the fitting procedure was
inappropriate. Here we shall use the model of \cite{monoclinic} and
take into account the thermal expansion strains properly. We shall
develop a consistent fitting procedure, free from the drawbacks of
the previous \cite{monoclinic} calculations, allowing us to obtain
the above mentioned unified description of high pressure effects,
thermal expansion, as well as physical properties of Rochelle salt
associated with the diagonal strains.

 Bulk Rochelle salt undergoes two second-order phase transitions at
$T_{\rm C1}=255$~K and $T_{\rm C2}=297$~K, with the intermediate
ferroelectric phase. Spontaneous polarization $P_1$ is directed
along the $a$ axis, accompanied by spontaneous shear strain
$\varepsilon_4$ in the $bc$ plane. The crystal is orthorhombic
(space group $P2_12_12_1$) in the paraelectric phases and
monoclinic ($P2_111$) in the ferroelectric phase.  As it follows from the
analysis of symmetry elements of its point group 222 and of
the uniaxial pressure point group $\infty/mmm$, no uniaxial or
biaxial pressure applied along the orthorhombic crystallographic
axes of a Rochelle salt crystal changes its symmetry. Neither do
the hydrostatic pressure or thermal expansion of the crystal.

The  mechanism of phase transitions in Rochelle salt remains
rather obscure. According to the most recent measurements
\cite{Suzuki,Petzelt3}, the largest displacements at a
ferroelectric phase transition are undergone by the O$_8$,
O$_{9}$, O$_{10}$ oxygens. It appears that the dipoles of the
Mitsui model, moving in the double-well potentials, should
plausibly be associated with the OH$_9$ and OH$_{10}$ groups.
Their motion, coupled with displacive vibrations of OH$_8$ groups
seems to be responsible for the phase transitions, as well as for
the spontaneous polarization. It would be interesting to elucidate
the pressure variation of the dipole moments, as well as of the
asymmetry parameters of the double-well potentials. This can shed
some light on the details of the transition mechanism.

The set of experimental data for Rochelle salt, used for
verification of the present modification of the Mitsui model,
includes the hydrostatic pressure dependences of transition
temperatures \cite{Bancroft,Samara} and dielectric permittivity
\cite{Samara}, the uniaxial \cite{Imai2,Unruh} and biaxial
\cite{biaxial} pressure dependences of the Curie temperatures.
Related to the diagonal strains,  components of piezoelectric (e.g.
$d_{1i}$, $i=1,2,3$) and elastic ($c_{i4}$) tensors
\cite{Schmidt,Fotchenkov,Sailer}, appearing in the ferroelectric
phase only, should be also described by the model. The other
characteristics, included into the fitting, are thermal expansion
coefficients and dilatations \cite{Bronowska,Imai1}, as well as
the diagonal-strain-related elastic constants\cite{Landolt}
$c_{ij}$ or compliances $s_{ij}$ ($i,j=1,2,3$).

We introduce the diagonal strains $\eps_{1}$, $\eps_2$, $\eps_3$
into the two-sublattice Mitsui model with the shear strain
$\eps_4$ \cite{ourrs} in the manner it was done in
\cite{monoclinic}. In section~2 we make a further modification of
the model by taking into account the host lattice contributions
into the thermal expansion. Section~3 contains the obtained
expressions for thermal, elastic, dielectric, and piezoelectric
characteristics. In section~4 we propose a new fitting procedure.
Using the found set of the fitting parameters for the Rochelle
salt crystals, we show that the developed theory is capable of describing the entire complex of the phenomena, related to the diagonal
strains: thermal expansion, temperature behavior of monoclinic
piezoelectric and elastic characteristics, hydrostatic, uniaxial,
and biaxial pressure dependences of the Curie temperatures in
Rochelle salt. The behavior of the dielectric permittivity under
hydrostatic pressure is described using the derived expression for
the permittivity of a partially clamped crystal.

\section{System thermodynamics in the presence of diagonal strains}
We consider an orthorhombic piezoelectric crystal in the paraelectric phase, to which an external hydrostatic  or
uniaxial or biaxial (along the crystallographic axes) pressure can
be applied. All these pressures produce  diagonal strains $\eps_i$
($i=1,2,3$); these strains are also present due to the thermal
expansion. Paraelectric piezoelectricity is associated with
the shear strain $\eps_4$. The Rochelle salt symmetry is presumed
having the spontaneous polarization directed along the $a$ axis and
coupled to the strain $\eps_4$.

We start with the modified two-sublattice Mitsui model with a
piezoelectric coupling to the shear strain $\eps_4$ and with the
diagonal strains \cite{ourrs,monoclinic}. A unit cell of the model
consists of two dipoles (two sublattices), oppositely oriented
along the $a$-axis (the axis of spontaneous polarization); they
are compensated in the paraelectric phases and get non-compensated
in the ferroelectric phase. The actual unit cell of a real
Rochelle salt crystal is twice as large and contains four dipoles.
At the transition to the monoclinic phase the angle between the
$b$ and $c$ axes changes from $\pi/2$ to $\pi/2-\eps_4$. The
diagonal strains $\eps_1$, $\eps_2$, $\eps_3$ describe the
relative changes in the lattice constants $a$, $b$, and $c$,
respectively, due to thermal expansion or under pressure.

 In the mean field
approximation, the model Hamiltonian reads \cite{ourrs}
 \begin{equation}
 \label{Hamiltonian}
  \hat H_{\mathrm{m}} = NU_{\mathrm{seed}}+\frac{N}{8}J(\eta_1^2+\eta_2^2)+\frac{N}{4}
K\eta_1\eta_2  -\sum\limits_q \Bigl[E(1)\frac{\sigma_{q1}}{2} -
E(2)\frac{\sigma_{q2}}{2}\Bigr],
\end{equation}
 where
$\eta_f\equiv\langle\sigma_{qf}\rangle$; $N$ is the number of the
unit cells;  $J$, $K$ are the Fourier-transforms (at ${\bf k}=0$)
of the constants of interaction between
 pseudospins belonging to the same and to different sublattices, respectively.

The phenomenological part of the Hamiltonian  $U_{\mathrm{seed}}$  is a
``seed'' energy of the host lattice of heavy ions which  forms the
asymmetric double-well potentials for the pseudospins. For the
case of Rochelle salt symmetry in presence of diagonal strains,
shear strain $\eps_4$, and field $E_1$, it reads
\begin{equation}
{U_{\mathrm{seed}}}=\frac{v_0}{2} c_{44}^{E0}\varepsilon_4^2
-{v_0}e_{14}^0\varepsilon_4  E_1 - \frac{v_0 \eps_0}{2}
\chi_{11}^{\varepsilon0} E_1^2 + \frac{v_0}{2}
\sum_{i,j=1}^3c_{ij}^{E0}\varepsilon_i\eps_j
-v_0\sum_{ij=1}^3c_{ij}^{E0}\alpha_i^0(T-T_i^0)\eps_j\,.
\end{equation}
Here $\eps_0$ is the vacuum permittivity; $v_0$ is the unit cell
volume of the model. The three first terms in $U_{\mathrm{seed}}$ are the
elastic, piezoelectric, and electric contributions due to the
shear strain $\eps_4$ and longitudinal electric field $E_1$. Two
last terms are related to the diagonal strains. Here
$c_{44}^{E0}$, $c_{ij}^{E0}$, $e_{14}^0$ are the ``seed''
constants describing the phenomenological contributions of the
crystal lattice into the corresponding observed quantities
$c_{44}^E$, $c_{ij}^E$, and $e_{14}$. In fact, the index $E$ in
$c_{ij}^{E0}$ is redundant, as the difference between the observed
$c_{ij}^{E}$ and $c_{ij}^{P}$ at $i,j=1,2,3$ is negligible. The
``seed'' quantities are zeros if the corresponding observed
quantities are zeros in the most symmetric phase (orthorhombic in
the case of Rochelle salt).

The last term, absent in the earlier model \cite{monoclinic}, is
the contribution of the host lattice into the energy of thermal
expansion. $\alpha_i^0$ are the ``seed'' thermal expansion
coefficients; $T_i^0$ are the temperatures at which the components
of this contribution vanish. It is known that the thermal strains
can be set to be equal to zero at any arbitrary temperature $T_0$
(the reference point for thermal expansion). This can be achieved
by choosing the values of $T_i^0$ accordingly; they will differ
from $T_0$ due to the pseudospin system contributions to the
thermal expansion.  To take into account  the ``seed''
contribution of the host lattice is indispensable for a proper
description of thermal expansion.

The coefficients
\begin{equation}
 E(1) = \frac12 J\eta_1 + \frac12 K\eta_2 + \Delta -
2\psi_4\varepsilon_4 + \mu_1E_1, \qquad \label{operator}  E(2) =
\frac12 J\eta_2 + \frac12 K\eta_1 - \Delta - 2\psi_4\varepsilon_4
+ \mu_1E_1
\end{equation}
in (\ref{Hamiltonian}) are the local mean fields acting on
pseudospins of the first and second sublattices in the $q^{\mathrm{th}}$ unit
cell. The parameter $\Delta$ describes the asymmetry of the double
well potential; $\mu_1$ is the effective dipole moment. The model
parameter $\psi_4$ describes the internal field created by the
piezoelectric coupling with $\varepsilon_4$ and essentially
determines the piezoelectric and elastic characteristics
associated with the shear strain $\eps_4$ \cite{ourrs,monoclinic}.
It is also assumed that a longitudinal electric field $E_1$ is
applied.
\begin{equation}\label{2.2}
    J \pm K= J_0 \pm K_0+2 \sum\limits_{i = 1}^3 {\psi_{i}^\pm \varepsilon _i
    }\,,\quad
\end{equation}
as well as the asymmetry parameter \begin{equation}
\label{2.2a}\Delta = \Delta _0 + \sum\limits_{i = 1}^3 {\psi_{3i}
\varepsilon _i }
\end{equation}
 are taken to be linear functions of the diagonal strains \cite{monoclinic}. Here
 $\psi_{3i}^\pm$ are introduced simply as the expansion
coefficients. However, for $J$ and $K$ such an expansion is
equivalent to taking into account the electrostrictive coupling
with the diagonal strains. For $\Delta$ it implicitly describes
the changes in the asymmetry parameter due to the changes produced by
external pressure or thermal expansion in the interatomic
distances and in the geometric parameters of the potential, like
the distance between the potential wells, etc. The parameters
$\psi_{3i}$, are analogous to the deformational potentials
$2\gamma=\partial \Delta/\partial \eps$ introduced in the
Anderson-Halperin-Varma-Phillips \cite{AHV,Phillips} model of
two-level systems with asymmetric double-well potentials in order
to describe ultrasound attenuation and thermal conductivity in
amorphous solids at very low temperatures.

The thermodynamic potential of the considered model is obtained
within the mean field approximation in the following form
 \begin{equation}
 \label{pot}
 g_{2E}(p_i,T) =U_{\mathrm{seed}} + \frac{J +K}4\xi^2 + \frac{J
-K}4\sigma^2
 - \frac{2\ln 2}{\beta} -  \frac1\beta\ln \cosh
\frac{\gamma + \delta}2
 \cosh \frac{\gamma - \delta}2+v_0\sum_{i=1}^4p_i\eps_i\,,
\end{equation}
where $\beta=1/k_{\mathrm{B}}T$, $k_{\mathrm{B}}$ is the Boltzmann constant, and
\[
\gamma = \beta\left( \frac{J + K}{2}\xi - 2\psi_4\varepsilon_4 +
\mu_1E_1\right), \qquad \delta = \beta \left( \frac{J - K}{2}\sigma
+ \Delta\right).
\]
Here $p_1=p_2=p_3=p_{\mathrm{h}}$ for hydrostatic pressure; $p_1\neq0$ and
$p_2=p_3=0$ for the uniaxial pressure applied along the axis $a$,
etc. The shear pressure $p_4$ is introduced formally, in order to
find the elastic and piezoelectric
characteristics associated with it; after that it is put equal to zero.

We introduced  the following linear combinations of the mean
pseudospin values
\[
\xi = \frac12(\eta_1 + \eta_2), \qquad \sigma = \frac12(\eta_1 -
\eta_2);
\]
$\xi$ is the parameter of ferroelectric ordering in the system.
The parameters $\xi$ and $\sigma$ are determined from the saddle
point of the thermodynamic potential (\ref{pot}): a minimum of
$g_{2E}$ with respect to $\xi$ and a maximum with respect to
$\sigma$ are realized at equilibrium. The corresponding equations
are
\begin{equation}
\label{ord-par} \xi = \frac{\sinh \gamma}{\cosh \gamma + \cosh
\delta}\,, \qquad \sigma = \frac{\sinh \delta}{\cosh \gamma + \cosh
\delta}\,.
\end{equation}

\section{Physical characteristics of Rochelle salt related to diagonal strains}

Using the following thermodynamic relations
\[
\frac{1}{v}\left( {\frac{\partial g_{2E}}{\partial \varepsilon _i
}} \right)_{E1} = 0,\qquad - \frac{1}{v}\left( {\frac{\partial
g_{2E}}{\partial E_1 }} \right)_\sigma = P_1 \qquad (i=1\div 4),
\]
where $v=v_0(1+\sum_{i=1}^3\eps_i)$ is the pressure and
temperature dependent unit cell volume, and retaining only linear
in $\eps_i$ terms in the ``seed'' contributions,
 we obtain expressions for strains and polarization
\begin{eqnarray}
&& \eps_i = -\sum\limits_{j = 1}^3 s_{ij}^{E0}
p_j +\alpha_i^0(T-T_i^0)
+\frac{1}{2v_0}\sum_{j=1}^3s_{ij}^{E0}(\psi_j^ + \xi ^2 + \psi_j^-
\sigma ^2
+ 2\psi_{3j} \sigma), \quad ({i } = 1\div 3), \label{strains123}\\
&& \varepsilon _4 = -\frac{p_4}{c_{44}^{E0}}+
\frac{e_{14}^0}{c_{44}^{E0}} E_1 - \frac{2\psi _4
}{v_0c_{44}^{E0}}\xi\, ,\label{strain4}\\[1.5ex]
 \label{2.4b} && P_1 = e_{14}^0 \varepsilon
_4 + \eps_0\chi _{11}^{\varepsilon 0} E_1 + \frac{\mu _1 }{v}\xi \,.
 \end{eqnarray}
 Here $s_{ij}^{E0}$ are the elements of the matrix inverse to the matrix of ``seed'' elastic constants $c_{ij}^{E0}$.

Two first terms in equation~(\ref{strains123}) represent the host
system contributions into the Hooke's law and thermal expansion
with regular pressure and temperature behavior. The sum in
equation~(\ref{strains123}) gives the pseudospin subsystem
contributions into the strains, having anomalous behavior in the
ferroelectric phase. The second term in equation~(\ref{strains123}) was
absent in the previous model \cite{monoclinic}.

From equations~(\ref{strains123})--(\ref{2.4b}) we can derive
expressions for other characteristics related to the diagonal strains. Thus, the coefficients of linear thermal
expansion are obtained in a rather cumbersome form
\begin{eqnarray}\alpha_{i}&=&\left(\frac{\partial
\eps_i}{\partial T}\right)_{\mathrm{p}}=\sum_{k=1}^3B_{ik}\Bigg\{\alpha_k^0+
\frac{1}{2v_0T(\varphi_2-\Lambda\varphi_3)}
\nonumber\\&&{}\times \sum_{j=1}^3 s_{kj}^{E0}\left[
\psi_j^+\xi(\lambda_2\delta-\varphi_3\gamma)+ ( \psi_j^- \sigma +
\psi_{3j})(\lambda_2\gamma-\tilde\varphi_5\delta)
\right]\Bigg\},
\end{eqnarray}
where \begin{eqnarray*}&& \hat B=\left[\hat I+\hat{s}^{E0}\hat{\tilde c}\right]^{-1},\\
&&\tilde c_{jk}=  - \frac{\beta}{2v_0(\varphi
_2-\Lambda\varphi_3)}\Bigl\{ (\psi_i^+ \varphi_{4j}+
\psi_j^+\varphi_{4i}) \xi -\psi_i^+\psi_j^+\xi^2\varphi_3+ (
\psi_i^- \sigma + \psi_{3i}  )( \psi_j^- \sigma + \psi_{3j}
)\tilde\varphi_5\Bigr\},\nonumber
\\
&& \tilde\varphi_5=  \lambda _1 - \left[ \frac{\beta (K +
J)}{4}+\Lambda\right](\lambda _1^2 - \lambda _2^2 );
\end{eqnarray*}
$\hat I$ is a unit matrix. The other notations are
\begin{eqnarray*}
&&\varphi _{4i} = \psi_i^+ \xi \varphi _3 - \left( {\psi_i^-
\sigma + \psi_{3i} } \right)\lambda _2 \,,\\[2ex]%
&&  \varphi _2 = 1 -
\frac{\beta J}{2}\lambda _1 - \beta ^2\frac{K^2 - J^2}{16}(\lambda
_1^2 - \lambda _2^2 ),\qquad\varphi _3 = \lambda _1 + \frac{\beta
(K - J)}{4}(\lambda _1^2 - \lambda _2^2
),\\
&& \lambda _1 = 1 - \xi ^2 - \sigma ^2, \qquad \lambda _2 = 2\xi
\sigma,\qquad \Lambda=\frac{2\beta\psi_4^2}{ v_0c_{44}^{E0}}\,.
 \end{eqnarray*}
 Alternatively, the coefficients of thermal expansion
can be found by numerical differentiation of
equation~(\ref{strains123}) for the strains $\eps_i$ with respect to
temperature; the results, of course, coincide. The coefficients
$\alpha_i$ are expected to have small anomalies at the Curie
temperatures \cite{Imai1,Bronowska}.

The molar specific heat at constant pressure of the model
 is obtained from the molar
entropy
\[
 S \!=\! -\frac{N_{\mathrm{A}}}{2} \Bigl(\frac{\partial
g_{1E}(\eps_i,T)}{\partial
T}\Bigr)_{\eps_i}\!\!=\frac{v_0N_{\mathrm{A}}}{2}\sum_{ij=1}^3c_{ij}^{E0}\alpha_i^0\eps_j
+\frac R2 \left( 2\ln 2 + \ln \cosh \frac{\gamma + \delta}2 \cosh
\frac{\gamma - \delta}2 - \gamma\xi - \delta\sigma\right)\!,
\]
where $g_{1E}(\eps_i,T)= g_{2E}(\sigma_i,T)-v_0\sum_i p_i\eps_i$;
$R$ is the universal gas constant, and $N_{\mathrm{A}}$ is the Avogadro
constant. Thus, the molar specific heat of the model is
\begin{eqnarray}
\label{cpheat} c_{\mathrm{p}}&=&T\left(\frac{\partial S}{\partial
T}\right)_{\mathrm{p}}=\frac{v_0N_{\mathrm{A}}T}{2}\sum_{ij=1}^3c_{ij}^{E0}\alpha_i^0\alpha_j+\frac{R}{4(\varphi_2-\Lambda\varphi_3)}\left\{
(\lambda_2\delta-\varphi_3\gamma)\left(\gamma+\xi\sum_{i=1}^3\frac{\alpha_i}{k_{\mathrm{B}}}\psi_i^+\right)\right.\nonumber\\
&&{}\left.- (\lambda_2\gamma-\tilde\varphi_5\delta)\left[\delta
-\sum_{i=1}^3\frac{\alpha_i}{k_{\mathrm{B}}}(\psi_i^- \sigma + \psi_{3i})\right]
\right\}.
\end{eqnarray}
As we shall see, it has small anomalies at the transition points. 
To obtain the total specific heat of a crystal that can be
compared to experimental data, we have to add to
equation~(\ref{cpheat}) a regular term linear in temperature (within the
considered temperature range) that would correspond to a
contribution of lattice vibrations not taken into account within
our model. Thus,
\begin{equation}
\label{totspecheat} c_{\mathrm{p}}^{\mathrm{tot}}=c_{\mathrm{p}}+c_{\mathrm{vibr}}\,,\qquad c_{\mathrm{vibr}} =A+ BT;
\end{equation}
the coefficients  $A$ and $B$ will be specified by fitting
equation~(\ref{totspecheat}) to experimental data. Often, an inverse
procedure is performed, when a regular linear contribution is
subtracted from the experimental data; the obtained result is then
compared with the theoretical specific heat of the ordering
subsystem.

The other found characteristics are, in particular, the elastic constants at constant electric
field ($i, j = 1,2,3$)
\begin{eqnarray}
 \label{2.6a}
 c_{ij}^E &=& - \Bigl(\frac{\partial p_i}{\partial
\eps_j}\Bigr)_{E,T}\nonumber\\
&=& c_{ij}^{E0} - \frac{\beta}{2v_0\varphi _2
}\Bigl[ (\psi_i^+ \varphi_{4j}+ \psi_j^+\varphi_{4i}) \xi
 -\psi_i^+\psi_j^+\xi^2\varphi_3+ (
\psi_i^- \sigma + \psi_{3i}  )( \psi_j^- \sigma + \psi_{3j}
)\varphi_5\Bigr],
\\
\label{2.6} c_{i4}^E &=& \frac{\beta \psi _4 }{v_0}\frac{\varphi _{4i}
}{\varphi _2},
\end{eqnarray}
as well as the monoclinic piezoelectric coefficients
\begin{eqnarray}
&& \nonumber   e_{1i} = \left(\frac{\partial P_1}{\partial
\eps_i}\right)_{E_1,T}=\frac{\mu _1 }{v}\left(\frac{\beta\varphi
_{4i} }{2\varphi _2 }-\frac{\xi}{1+\sum_{i=1}^3\eps_i}\right),
\\\label{2.7} && d_{1i} = \left(\frac{\partial \eps_i}{\partial
E_1}\right)_{p_i,T}= \sum\limits_{j = 1}^4 {s_{ij}^E e_{1j} }\,,
\end{eqnarray}
where $s_{ij}^E$ is the matrix of elastic compliances, inverse to
the matrix of elastic constants $c_{ij}^E $, and
\begin{eqnarray*}
&&\varphi _5 = \lambda _1 - \frac{\beta (K + J)}{4}\left(\lambda _1^2 -
\lambda _2^2 \right).
\end{eqnarray*}
The other piezoelectric and elastic characteristics are
\[ h_{1i} = - \left( \frac{\partial E_1}{\partial \varepsilon_i}
\right)_{P_1} =
 \frac{e_{1i} }{\eps_0\chi _{11}^\varepsilon }\,,\qquad
 g_{1i} = - \left( \frac{\partial E_1}{\partial p_i}
\right)_{P_1}= \frac{d_{1i} }{\eps_0\chi _{11}^\sigma }\,, \qquad
c_{ij}^P =\left(\frac{\partial p_i}{\partial \eps_j}\right)_{P,T}=
c_{ij}^E + e_{1i} h_{1j}\,.
\]
Here,
 \begin{equation}
 \label{2.43}
\chi_{11}^{\varepsilon} = \frac{1}{\eps_0}\left(\frac{\partial
P_1}{\partial E_1}\right)_\eps=\chi_{11}^{\varepsilon 0} +
  \frac{\beta\mu_1^2}{2v\eps_0}
  \frac{\varphi_3}{\varphi_2}
  \end{equation}
is the dielectric susceptibility of a clamped crystal, and
 \begin{equation}
 \label{2.42}
\chi _{11}^\sigma = \frac1{\eps_0}\left(\frac{\partial
P_1}{\partial E_1}\right)_{\mathrm{p}}= \chi_{11}^{\sigma 0}+
 \frac{\beta(\mu_1')^2}{2v\eps_0}\frac{\displaystyle \varphi_3}
{\displaystyle\varphi_2- \Lambda \varphi_3} +
\frac1{\eps_0}\sum\limits_{i = 1}^3 {e_{1i} d_{1i} }
\end{equation}
is the static dielectric susceptibility of a mechanically free
crystal \cite{monoclinic}. Here we introduce the following
notations
\[ \quad {\mu_1'} = {\mu_1} - {2} \psi_4 d_{14}^0\,,\qquad
  d_{14}^0 = \frac{ e_{14}^0}{ c_{44}^{E0}}\,,\qquad
\chi_{11}^{\sigma 0} = \chi_{11}^{\varepsilon 0} +
\frac1{\eps_0}e_{14}^0d_{14}^0\,.
\]
In paraelectric phases, this expression for the free
susceptibility coincides with that obtained within the modified
Mitsui model without thermal strains \cite{ourrs2}. The sum
${\eps_0^{-1}}\sum e_{1i}d_{1i}\,$, different from zero in the ferroelectric
phase and not exceeding 5\% of the total susceptibility,  was
absent in the earlier model.

As one can easily verify, monoclinic quantities $e_{1i} $,
$d_{1i} $, $h_{1i} $, $g_{1i} $, $c_{i4}^E $, $c_{i4}^P $ ($i =
1,2,3)$ differ from zero only at non-zero polarization, in
agreement with the symmetry considerations.

The temperature of the second order phase transition is determined
from the condition that dielectric susceptibility of a free
crystal $\chi_{11}^{\sigma}$ diverges at $T \to T_{\rm C}$. From
equation~(\ref{2.42}) and using equation~(\ref{ord-par}) we obtain
\begin{equation}
\label{tc} \cosh^2\left(  \frac{ J -  K}{4k_{\mathrm{B}}T_{\rm C}}\sigma_{\mathrm{c}} +
\frac{\Delta}{2k_{\mathrm{B}}T_{\rm C}}\right) = \frac{ K +  J}{4k_{\mathrm{B}}T_{\rm
C}} + \frac{2 \psi_4^2 }{v_0 c_{44}^{E0}k_{\mathrm{B}}T_{\rm C}}\,,
\end{equation}
where the model parameters  $J$, $K$, $\Delta$ are taken at
$T_{\rm C}$, being renormalized by diagonal strains according to
equation~(\ref{2.2}).

Equation (\ref{tc}) is valid both for the ambient pressure case
and for the stressed crystal. It can be rewritten in the two
following convenient forms:
\begin{equation}
\label{tc2} \sigma_{\mathrm{c}}=\sqrt{1-\frac{k_{\mathrm{B}}T_{\rm C}}{\frac{ K +
J}{4} + \frac{2\psi_4^2 }{v_0 c_{44}^{E0}}}}\,,
\end{equation}
which gives an explicit expression for $\sigma$ at the transition
points, and
\begin{equation}
\label{tc3} \sum_i \delta _{3i}\eps_{ci} +\sigma_{\mathrm{c}}\sum_i
\psi_i^-\eps_{ci} =-\Delta_0-\frac{J_0-K_0}{2}\sigma_{\mathrm{c}} +
2k_{\mathrm{B}}T_{\rm C}{\rm Arccosh}{\sqrt{\frac{ K + J}{4k_{\mathrm{B}}T_{\rm C}} +
\frac{2 \psi_4^2 }{v_0 c_{44}^{E0}k_{\mathrm{B}}T_{\rm C}}}}\,,\nonumber
\end{equation}
useful in the fitting procedure. Here $\eps_{ci}$ are the strains
at the Curie temperature.

\section{Numerical calculations}

\subsection{Fitting procedure}

The model parameters must provide a fit of the theory to the
experimental data for the following characteristics: the Curie
temperatures at ambient pressure $T_{{\rm C}k}$ ($k=1,2$ in
Rochelle salt) and their hydrostatic and uniaxial pressure slopes
$\partial T_{{\rm C}k}/\partial p_{\mathrm{h}}$ and $\partial T_{{\rm
C}k}/\partial p_j$, the temperature curves of thermal expansion
strains $\eps_i$, linear thermal expansion coefficients,
monoclinic piezomodules $g_{1i}$, and elastic constants $c_{ij}$
and $c_{i4}$ ($i,j=1,3$). Simultaneously we check for an agreement
with experiment for the quantities previously described
\cite{ourrs,ourrs2} by the modified Mitsui model without thermal
strains, such as spontaneous polarization, static free and clamped
dielectric susceptibilities $\chi_{11}^{\sigma,\varepsilon}$,
piezomodule $d_{14}$, specific heat, elastic constant at constant
field $c_{44}^E$, as well as microwave dielectric permittivity
$\eps_{11}(\nu,T)$. A detailed analysis of the effect of diagonal
strains on the physical characteristics of Rochelle salt
associated with the shear strain $\eps_4$ will be given elsewhere.

The adopted values of the model parameters are given in
table~\ref{table-parameters}; details of the fitting procedure are
described below.
\begin{table}[hbt]
\vspace{-2mm}
\caption{The model parameters used for description of Rochelle
salt. }\label{table-parameters}
\begin{center}
\small
\begin{tabular}{cccccccccccccc}
\hline\hline
$\vphantom{\frac{1^1}{2^2}}$
  $\bar a_0$ & $\bar b_0$ &  $J_0/k_{\mathrm{B}}$&  $K_0/k_{\mathrm{B}}$ & $\Delta_0/k_{\mathrm{B}}$ & $\psi_4/k_{\mathrm{B}}$  & $\mu_1^0$& $k_{\mathrm{T}}$ & $e_{14}^0$ & $ \chi_{11}^{\sigma0}$\\
  & &   \multicolumn{4}{c}{(K)} & ($10^{-30}$~C$\cdot$m)  & (K$^{-1})$  & C/m$^2$\\
 \hline
   0.3162 & 0.662 & 764.63 &  1476.46 & 745.14 & $-750$
 & $8.7$ & $-0.0008$&   0.033 &10.1 \\[2ex]
 \hline\hline
$\vphantom{\frac{1^1}{2^2}}$  $\psi_1^+/k_{\mathrm{B}}$&  $\psi_2^+/k_{\mathrm{B}}$ & $\psi_3^+/k_{\mathrm{B}}$ & $\psi_1^-/k_{\mathrm{B}}$&  $\psi_2^-/k_{\mathrm{B}}$ & $\psi_3^-/k_{\mathrm{B}}$& $\psi_{31}/k_{\mathrm{B}}$&  $\psi_{32}/k_{\mathrm{B}}$ & $\psi_{33}/k_{\mathrm{B}}$\\
   \multicolumn{9}{c}{(K)} \\
\hline
  $-13000$ & $-12000$ & $-9850$ & $10614$ & $13537$ & $1080$ & $-12125$ &$-15993$ & $-6043$\\[2ex]
 \hline\hline
$\vphantom{\frac{1^1}{2^2}}$ $\alpha_1^0$&  $\alpha_2^0$ & $\alpha_3^0$ & $T_1^0$&  $T_2^0$ & $T_3^0$
\\
 \multicolumn{3}{c}{($10^{-5}$ K$^{-1}$)}& \multicolumn{3}{c}{(K)} \\
\hline
   5.800& 3.353& 4.333&   238.35 & 185.32 & 311.71\\[2ex]
\hline\hline
$\vphantom{\frac{1^1}{2^2}}$  $c_{11}^{E0}$&  $c_{12}^{E0}$ & $c_{13}^{E0}$ & $c_{22}^{E0}$ & $c_{23}^{E0}$ & $c_{33}^{E0}$ & $c_{44}^{E0}$\\
 \multicolumn{7}{c}{($10^{10}$ N/m$^{2}$)} & \\\hline
  2.842 & 1.794 & 1.541 & 4.219 & 2.031 & 3.987 & 1.18  &
 \\\hline\hline
\end{tabular}
  \vspace{-3mm}
\end{center}
\end{table}

\normalsize
The temperature variation of the order parameter $\xi$ is
determined by numerical minimization of the thermodynamic
potential (\ref{pot}); $\sigma$ is found from equation~(\ref{ord-par});
the strains are determined from equations~(\ref{strains123}) and
(\ref{strain4}). As the reference point for thermal expansion
(where $\eps_i=0$ at $p_i=0$) we chose the upper transition
temperature $T_{\rm C2}=297$~K. This condition allows us to
express $T^0_i$ from equation~(\ref{strains123}) via $\alpha_i^0$,
$\psi_{i}^-$, and $\psi_{3i}$. Strictly speaking, $T^0_i$ are not
the fitting parameters of the model, since the reference point can be
chosen arbitrarily. At 308~K and at ambient pressure, the lattice
constants are \cite{9} $a=11.927$~\AA, $b=14.292$~\AA,
$c=6.225$~\AA.

The ``seed'' linear thermal expansion coefficients $\alpha_{i}^0$
were chosen to yield the thermal strains $\eps_i$ at $T_{\rm C1}$
equal to $\alpha_{i}^{275}(T_{\rm C1}-T_{\rm C2})$, where
$\alpha_{i}^{275}$ are the experimental \cite{Imai1} values of the
expansion coefficients in the middle of the ferroelectric phase
(at 275~K).

 Special care has
been taken that below 20~kbar for the hydrostatic pressure and
below 200~bar for uniaxial or biaxial pressures and between 0 and
350~K, no additional  phase transition takes
place in the system, apart from those taking place at ambient pressure. 

The number and (if any) temperature and order of the phase
transitions for the Mitsui model without thermal strains are
usually analyzed in terms of the dimensionless variables $\bar a$
and $\bar b$
\begin{equation}
 \bar a = \frac{ K -  J}{K +  J +
\frac{8}{v_0}\psi_4^2s_{44}^{E0}}\,, \qquad  \label{dimensionless}
\bar b = \frac{8 \Delta}{ K + J + \frac{8}{
v_0}\psi_4^2s_{44}^{E0}}
 \end{equation}
 and the dimensionless  transition temperature
 \begin{equation}\label{tbar} \bar t^0_{\mathrm{c}} =
\frac{4k_{\mathrm{B}}T_{\rm C}}{ K +  J + \frac{8}{ v_0}\psi_4^2s_{44}^{E0}}\,.
\end{equation}
 The phase diagram of the
conventional (undeformable) Mitsui model in the  $(\bar a, \bar
b)$ plane \cite{vaks,werch,Dublenych} shows the regions with
different numbers and types of the phase transitions; its topology
is not changed by inclusion of the shear strain $\eps_4$. It has
been found that only in a very narrow region of the $(\bar a, \bar
b)$ plane, the system undergoes two second order phase transitions
with the intermediate ferroelectric phase.

In the presence of diagonal strains, $\bar a$ and $\bar b$ become
functions of temperature and pressure. In the fitting procedure we
shall deal with the values of $\bar a$ and $\bar b$ at the upper
Curie temperature and at ambient pressure $\bar a_0$ and $\bar b_0$.
Absence of additional phase transitions at the chosen values of
the model parameters is verified directly, by calculating the
order parameter at all temperatures between 0 and 350 K and at
pressures below 25~kbar (hydrostatic) or 200~bar (uniaxial). The
values of $\bar a_0$ and $\bar b_0$ should be from the same region
of the $(\bar a, \bar b)$ phase diagram of the undeformable Mitsui
model that yield two second order phase transitions.
  With decreasing $\bar a_0$, the maximal values of spontaneous
polarization, spontaneous strain $\eps_4$, and anomalous parts of
diagonal strains increase. We choose the value of $\bar a_0$ that
gives the best agreement with experiment for these
characteristics. Once the values of $\bar a_0$, $\bar b_0$,
$\psi_4$, and $c_{44}^{E0}$ are chosen, we are in a position to
find $J_0$, $K_0$, and $\Delta_0$, using equations~(\ref{tc2}),
(\ref{dimensionless}), and (\ref{tbar}).

The parameters $c_{44}^{E0}$, $\psi_4$, and $\mu_1$ are varied
around their values obtained in the previous study \cite{ourrs} in order
to get the best fit for spontaneous polarization $P_1$,
piezoelectric coefficient $d_{14}$, static free and clamped
$\chi_{11}^{\sigma,\eps}$ dielectric susceptibilities and dynamic
dielectric permittivity $\eps_{11}(\nu,T)$. The dipole moment
$\mu_1$ is assumed to decrease linearly with an increasing temperature
as
\[\mu_1=\mu_1^0\left[1+k_{\mathrm{T}}(T-T_{\rm C2})\right].\]
The values of $\mu_1^0$ and $k_{\mathrm{T}}$ are given in
table~\ref{table-parameters}.

We require that the theoretical values of the elastic constants
$c_{ij}^E$ ($i,j=1,3$) at $T_{\rm C2}$  should coincide with their
experimental values \cite{64B2}, available for 307~K (this is a
reasonable approximation due to a very weak temperature
dependence of $c_{ij}^E$). Thus, we can easily determine
$c_{ij}^{E0}$ using equation~(\ref{2.6a}).

It is required that the best possible description of
experimental \cite{Imai2,Unruh} uniaxial pressures dependence of
the two Curie temperatures should be obtained. To this end, at the chosen $a_0$,
$b_0$, $\psi_4$, $c_{44}^{E0}$, $\psi^+_i$, the six parameters
$\psi^-_i$ and $\psi_{3i}$ are determined from six linear
equations (\ref{tc3}) written at $T_{{\rm C}k}^i=T_{{\rm
C}k}^0+(\partial T_{{\rm C}k}/\partial p_i)p_i$ ($k=1,2$ and
$i=1,2,3$) at $p_i=100$~bar and combined with equation~(\ref{tc2}). The
slopes $\partial T_{{\rm C}k}/\partial p_i$ were varied around
their experimental
values\cite{Imai2}. 
 The strains at the Curie
temperatures were approximated as
\[
\eps_{cl}(T_{{\rm C}k}^i)=\alpha_l^{275}(T_{{\rm C}k}^i-T_{\rm
C2}^0)-s_{li}^{E0}p_i
\]
during the fitting. The obtained values of $\delta^-_i$ and
$\psi_{3i}$  were found to be independent of the used values of
$p_i$.

One of $\psi_i^+$ parameters, say, $\psi_3^+$, can be determined
from the condition that the lower Curie temperature should be
$T_{{\rm C}1}=255$~K. For the  two remaining parameters $\psi_1^+$
and $\psi_2^+$ there is the condition that the two calculated
transition temperatures at hydrostatic pressure of 1~kbar and the
lower transition temperature at 20~kbar would be in agreement with
the experimental data \cite{Bancroft,Samara}. However, the
dependences of the transition temperatures on uniaxial pressures
and on the hydrostatic pressure are not completely independent,
(this will  be discussed later). Therefore, normally $\psi_1^+$
and $\psi_2^+$ can be varied continuously in certain ranges and
provide correct theoretical dependences $T_{{\rm C}k}(p_{\mathrm{h}})$ at
given $\bar a_0$, $\bar b_0$ and $\psi_4$. From these ranges we
should select the values which yield the best fit for the
temperature curves of the piezoelectric coefficients $g_{1i}$,
anomalous parts of thermal strains $\eps_{si}$ in the
ferroelectric phase, and thermal expansion coefficients
$\alpha_i$. It should be mentioned, however, that a perfect fit
both for $g_{1i}$ (\cite{Schmidt}) and $\eps_{si}$ (\cite{Imai1})
 cannot be obtained simultaneously, so a certain compromise has to
be made.

A criterion to make an unambiguous choice of the theory parameters
hardly exists. Since the chosen set of $\psi_{3i}^\pm$ is not unique,
it is not possible to precisely establish the temperature and pressure
variation of the interaction constants. However, the
overall tendency is such that the hydrostatic compression enhances
the asymmetry parameter $\Delta$, as well as the constants of
interactions between the pseudospins within the same and in
different sublattices. Pressure slopes of $\Delta$, $J$, and $K$
are very sensitive to the choice of $\bar b_0$ at given $\bar
a_0$, while the observable quantities are not that much sensitive. The average
slopes are about 1--3\%/kbar  for $J$ and 4.5--7\%/kbar for $\Delta$
and $K$. This issue will be explored in more detail elsewhere.

\subsection{Thermal expansion and specific heat}
\begin{figure}[ht]
\includegraphics[width=0.46\textwidth]{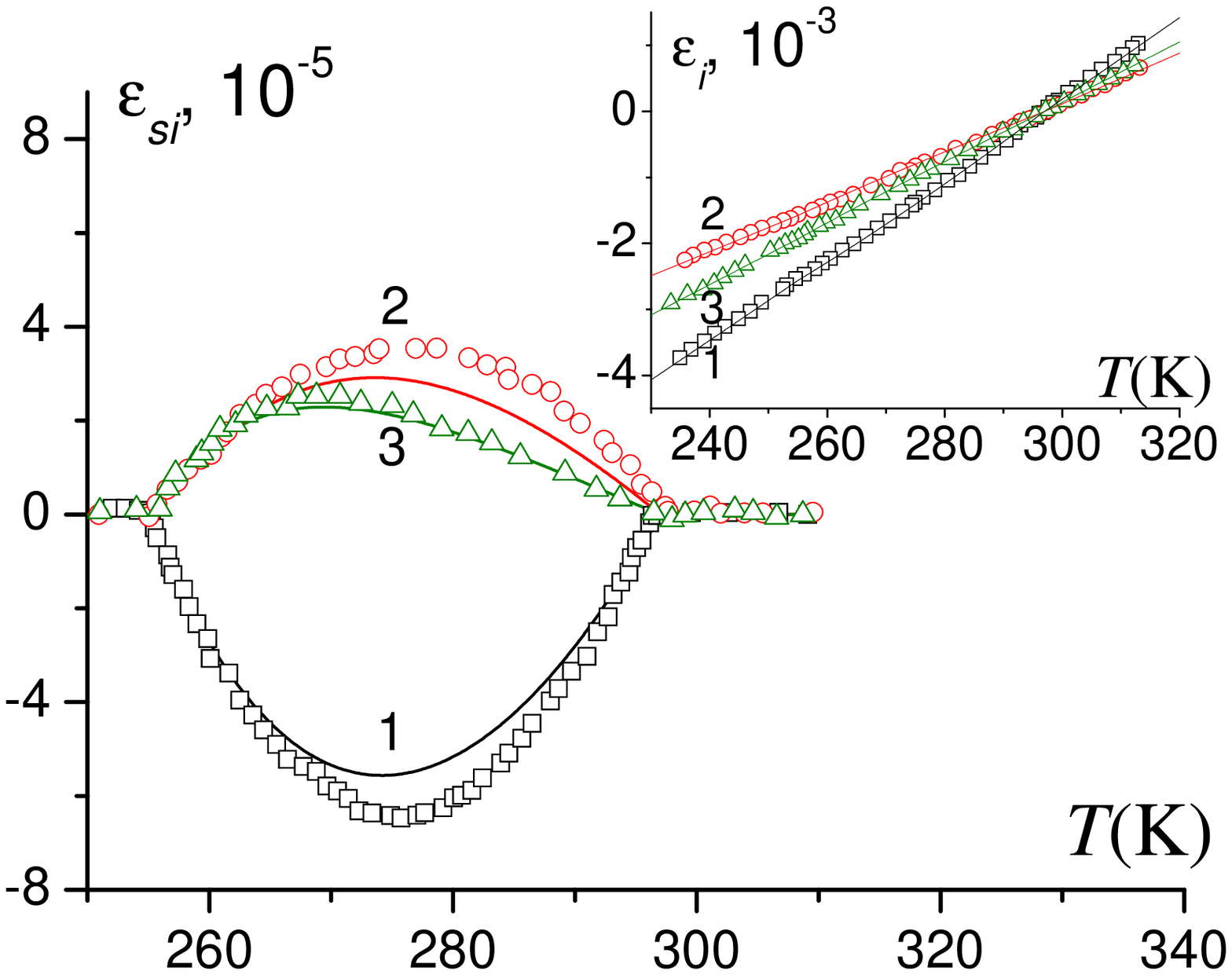}%
\hfill%
\includegraphics[width=0.46\textwidth]{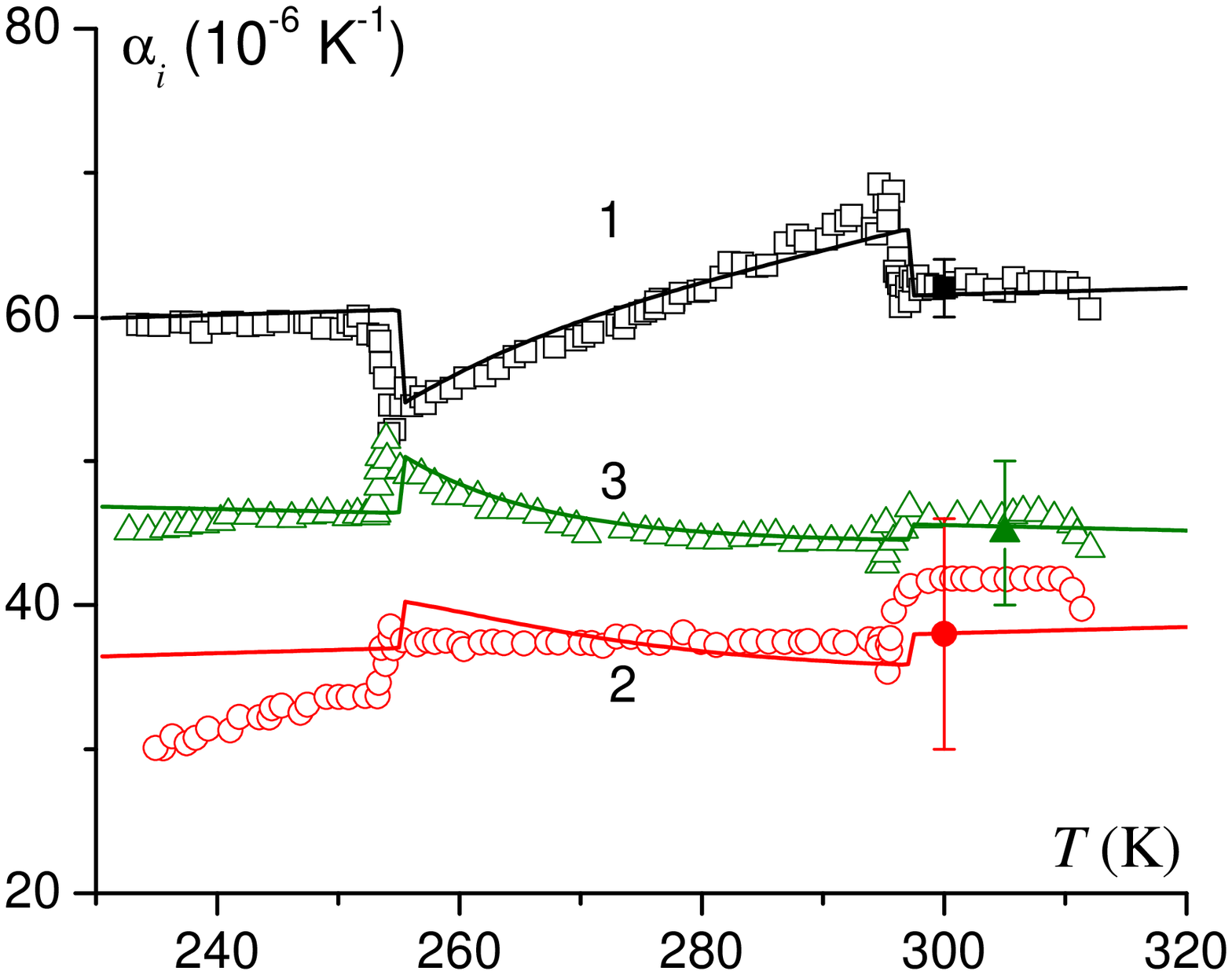}%
\\%
\parbox[t]{0.48\textwidth}{%
\caption{\label{fig_strains}  Anomalous parts of thermal diagonal
strains of Rochelle salt as functions of temperature: 1,
$\square$: $\eps_{s1}$; 2, $\bigcirc$: $\eps_{s2}$; 3,
$\bigtriangleup$: $\eps_{s3}$. Inset: total strains as functions
of temperature: 1, $\square$: $\eps_{1}$; 2, $\bigcirc$:
$\eps_{2}$; 3, $\bigtriangleup$: $\eps_{3}$ Lines: a theory;
symbols: experimental points taken from \cite{Imai1}.}%
}
\hfill%
\parbox[t]{0.48\textwidth}{%
\caption{\label{fig-deriv} Coefficients of linear thermal
expansion of Rochelle salt as functions of temperature. 1,
$\square$, $\blacksquare$: $\alpha_{1}$; 2, $\bigcirc$, $\bullet$:
$\alpha_{2}$; 3, $\bigtriangleup$, $\blacktriangle$: $\alpha_{3}$.
 Lines: a theory. Open and closed symbols are experimental points
taken from \cite{Imai1} and from \cite{dosSantos}, respectively.}%
}
\end{figure}

The temperature dependence of the diagonal strains $\eps_i$ caused
by thermal expansion of a crystal in the absence of external pressures
is plotted in the inset to figure~\ref{fig_strains}.

The experimental points, obtained from the data for thermal
dilatations \cite{Imai1}, are well described by the proposed theory.

In the ferroelectric phase, the $\eps_i(T)$ curves have small
bucklings (anomalous parts) caused by electrostrictive
coupling to spontaneous polarization. In order to extract these
anomalous parts of the diagonal strains, Imai \cite{Imai1}
extrapolated the measured temperature curves of the strains from
the paraelectric phases onto the ferroelectric phase and
subtracted them from the total measured strains. With the same
purpose, we calculate some hypothetical paraelectric strains
$\eps_i^{\mathrm{h}}$ (coinciding in the paraelectric phases with the actual
strains $\eps_i$) from equation~(\ref{strains123}) by putting $\xi=0$
at all temperatures and determining $\sigma$ from
equation~(\ref{ord-par}), and find the spontaneous strains as
$\eps_{si}=\eps_i-\eps_i^{\mathrm{h}}$. The obtained results are shown in the
major part of figure~\ref{fig_strains}. As one can see, at the
adopted values of the model parameters, the theory well reproduces
the asymmetric shape of the $\eps_{s3}(T)$ curve, but
underestimates the magnitude of $\eps_{s1}$ and $\eps_{s2}$. The
agreement can be improved by choosing different values of the
model parameters, but the agreement with $g_{1i}(T)$ will be
spoiled.

The corresponding linear thermal expansion coefficients are shown
in figure~\ref{fig-deriv}. The theoretical and experimental values
of their jumps at the Curie temperatures  are summarized in
table~\ref{table-jumps}.
\begin{table}[h]
  \caption{The calculated jumps of the thermal expansion coefficients and specific heat at Curie temperatures.
  The values in parentheses are experimental data of \cite{Imai1}. }\label{table-jumps}
  \vspace{1ex}
 \begin{center} \begin{tabular}{ccccc}\hline\hline
 & $\Delta \alpha_{1}$  &
$\Delta \alpha_{2}$ & $\Delta \alpha_{3}$ & $\Delta c_{\mathrm{p}}$  \\
 & \multicolumn{3}{c}{($10^{-6}$K$^{-1})$} & {(J/mol K)} \\
\hline
$T_{\rm C1}$ & --6.0 (--7.0)& 3.0 (3.5)& 3.7 (4.5)& 0.74 \\
$T_{\rm C2}$ & 4.7 (7.1) &--2.1 (--3.7)& --1.1 (--2.0)& 1.09 \\
\hline\hline
  \end{tabular}
  \vspace{-2mm}
  \end{center}
  \vspace{-3mm}
\end{table}

We have a fairly good agreement with experiment \cite{Imai1} for
$\alpha_{1}$ and $\alpha_3$, both for the signs and for the values
of the coefficient anomalies at the transition points, as well as
for the temperature slope of the curves in the ferroelectric
phase. The agreement is worse for $\alpha_2$. It should be noted
that the experimental  behavior of $\alpha_2(T)$ is somewhat
different from that of $\alpha_{1}$ and $\alpha_3$, with a marked
decrease below the lower Curie temperature and a large difference
between  the coefficient values in the two paraelectric phases. No
perceptible temperature variation in the ferroelectric phase was
experimentally detected, also in contrast with the $\alpha_{1}$
and $\alpha_3$ behavior. The theoretical temperature curve of
$\alpha_2(T)$ is very much like that of $\alpha_3$ and
qualitatively similar to that of $\alpha_{1}$. The theoretical
$\alpha_2$ in the upper paraelectric phase well accords with the
single experimental value of \cite{dosSantos} obtained from
the synchrotron radiation Renninger scan, although the reported
error of these measurements is so large that the values of
\cite{Imai1} also fall in this error range (see
figure~\ref{fig-deriv}).

\begin{figure}[ht]
\centerline{\includegraphics[width=8cm]{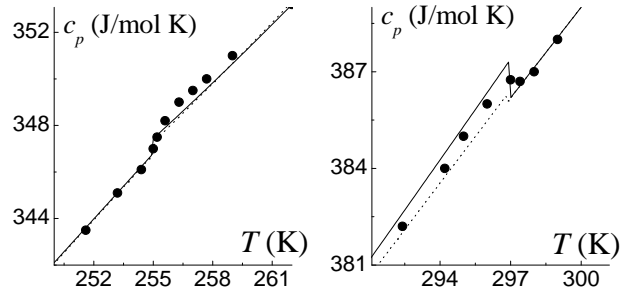}}
\caption{\label{fig-specheat}Specific heat of Rochelle salt as a
function of temperature. Lines: a theory. Solid line: this work;
dashed line: the modified Mitsui model without the thermal strains
\cite{ourrs}. Symbols: experimental points taken from
\cite{Tatsumi}.  }
\end{figure}
Figure~\ref{fig-specheat} shows that the present model yields a
better agreement with experimental data for the small anomalies of
specific heat of Rochelle salt at the Curie points than it was
obtained with the earlier model \cite{ourrs}, especially for the
magnitude of the upper anomaly. The regular contribution of
lattice vibrations was taken to be $c_{\mathrm{vibr}}=105.845+0.855T$
(J/mol K) for the present model and $103.456+0.944T$ (J/mol K) for
the previous model.

The calculated values of the specific heat jumps are given in
table~\ref{table-jumps}. We obtain positive anomalies at both
transition points, in accordance with the most recent measurements
\cite{Tatsumi}. The jumps were defined as the differences between
the ferroelectric and paraelectric values of specific heat or
expansion coefficients at the Curie points.

\subsection{Diagonal-strain-related piezoelectric and elastic
constants}

\begin{figure}[htb]
\includegraphics[width=0.46\textwidth]{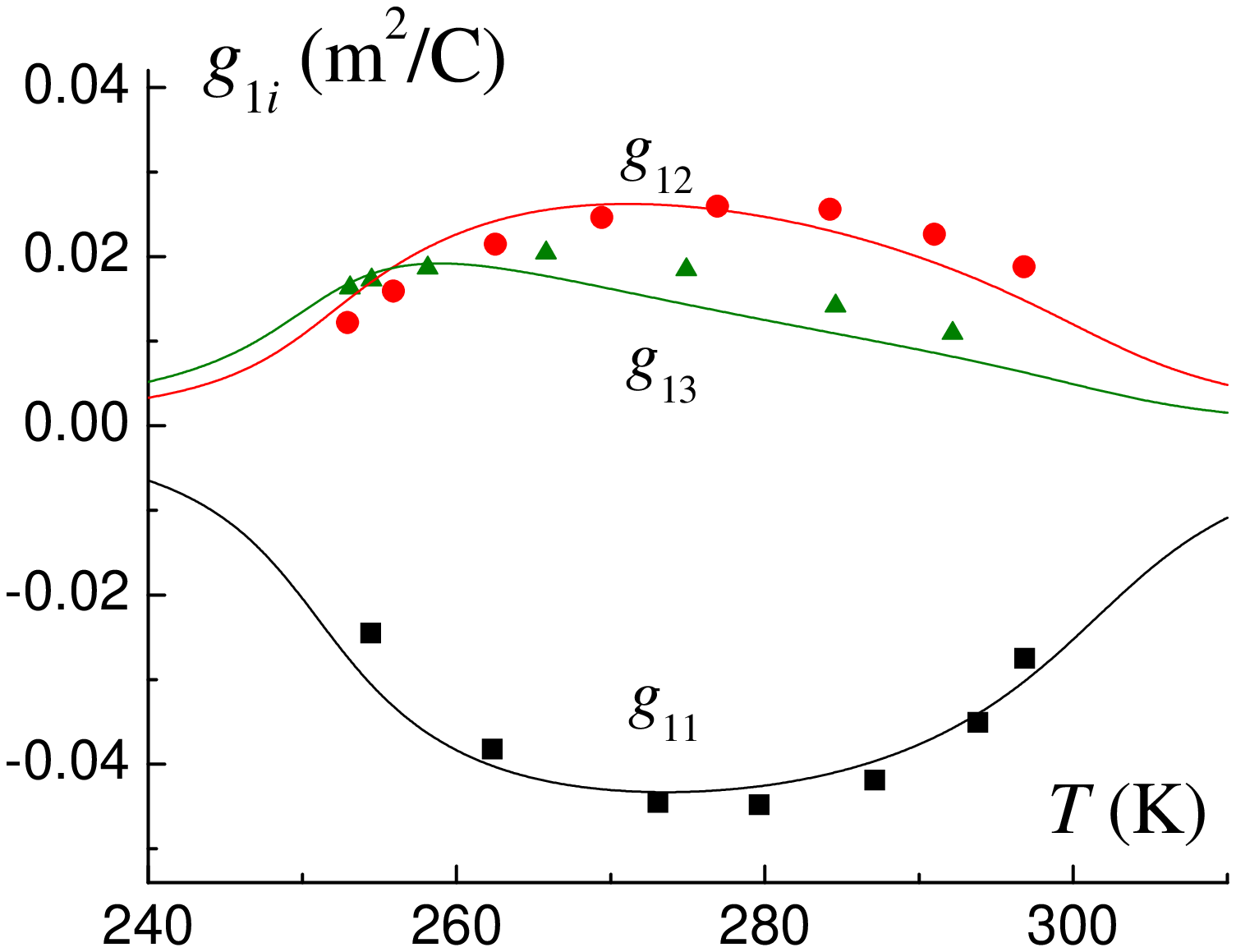}%
\hfill%
\includegraphics[width=0.46\textwidth]{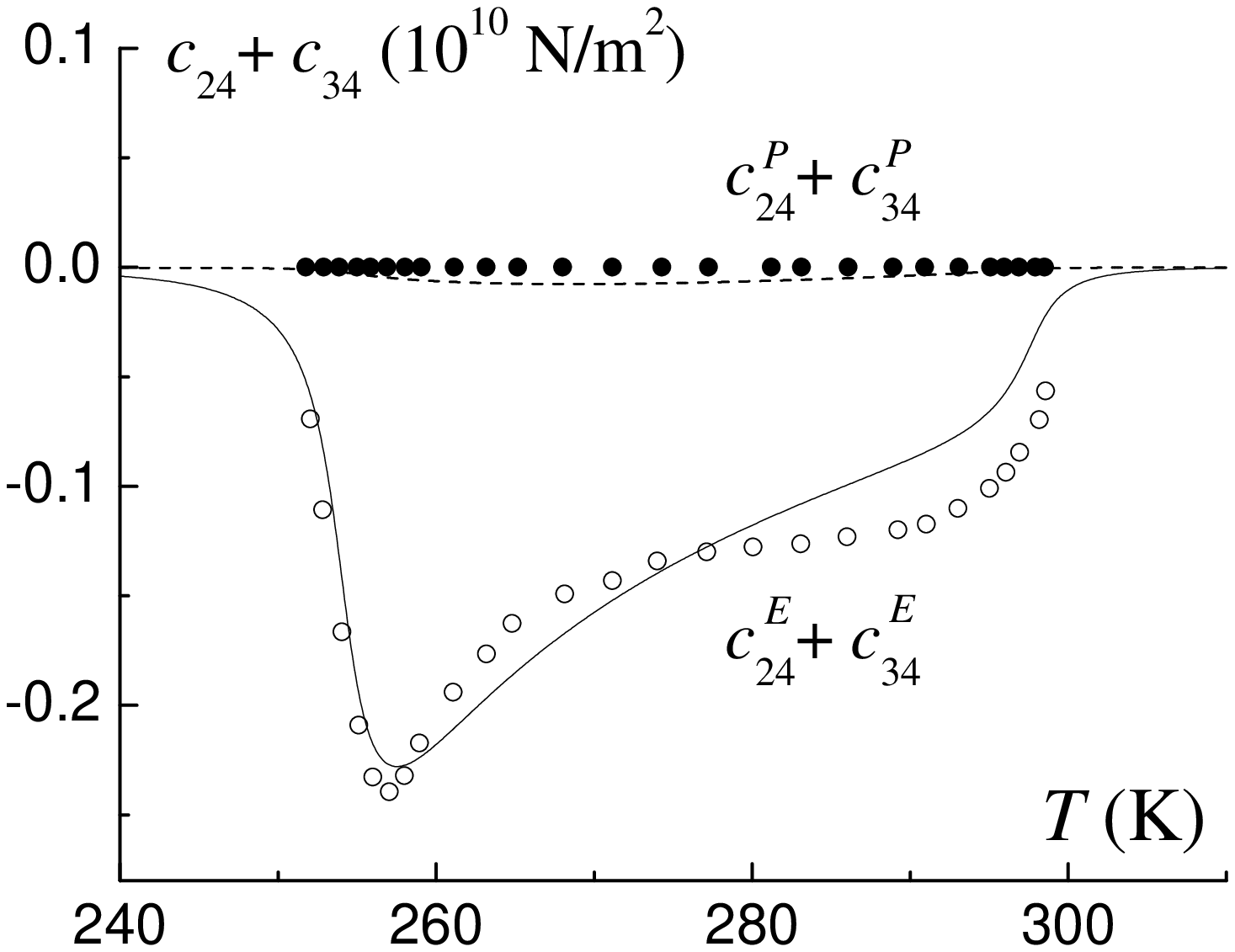}%
\\%
\parbox[t]{0.48\textwidth}{%
\caption{\label{fig-g1i}  \looseness=-1Piezoelectric constants of Ro\-chelle salt
as functions of temperature at $E_1=5$~kV/cm: 1, $\square$:
$g_{11}$; 2, $\bigcirc$: $g_{12}$; 3, $\bigtriangleup$: $g_{13}$.
Lines: a theory; symbols: experimental points taken from
\cite{Schmidt}.}%
}
\hfill%
\parbox[t]{0.48\textwidth}{%
\caption{\label{fig-c24c34} Elastic constants
$c_{24}^{E}+c_{34}^E$ and $c_{24}^{P}+c_{34}^P$ of Rochelle salt
as functions of temperature at $E_1=500$~V/cm. Lines: a theory;
symbols: experimental points taken from \cite{Sailer}.}%
}
\end{figure}

Figure~\ref{fig-g1i} shows temperature dependences of
piezoelectric  constants $g_{1i}$.
As one can see, a fairly good
agreement with experiment is obtained, including the opposite
signs of $g_{11}$ and
of $g_{12}$ and $g_{13}$, asymmetric shape
of $g_{13}(T)$ dependence, as well as the interception of the
$g_{12}$ and $g_{13}$ curves
near the lower Curie point. The
overall behavior of $g_{1i}(T)$ is similar to that of
$\eps_{si}(T)$.

We do not depict the calculated elastic constants $c_{ij}^E$ and
$c_{ij}^P$ ($i,j=1,2,3$), as they are practically temperature
independent between 230 and 330 K. A very small variation can be
detected in the ferroelectric phase for  $c_{ij}^E$, with the
difference between $c_{ij}^E$ and  constant $c_{ij}^P$ being less
than 1\% of $c_{ij}^E$ at most. The maximal values of
monoclinic constants $c^E_{i4}$ are more than by one order of
magnitude smaller than $c_{ij}^E$. The shape of the experimental
$c_{24}^{E}+c_{34}^E$ vs $T$ curve is qualitatively reproduced by
the proposed model, as seen in figure~\ref{fig-c24c34}. A
quantitative agreement with experiment is reasonable.

\begin{figure}[ht]
\centerline{\includegraphics[width=0.46\columnwidth]{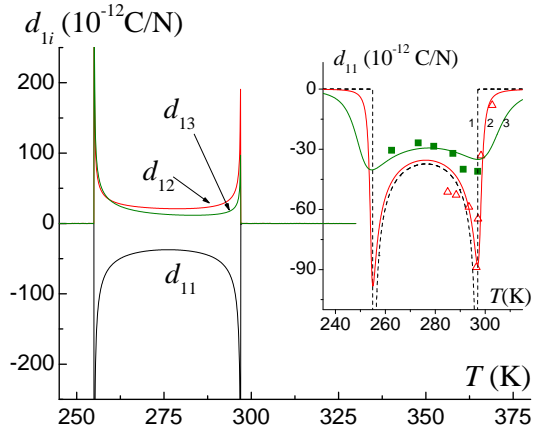}}
\caption{\label{fig-d1i}  Piezoelectric coefficients $d_{1i}$ of
Rochelle salt as functions of temperature. The inset: $d_{11}$ as
a function of temperature at different fields $E_1$ (kV/cm): 1, 0;
2, $\bigtriangleup$: 0.304; 3, $\blacksquare$: 5. Lines: a theory;
symbols: experimental points taken from \cite{Schmidt}
($\blacksquare$) and \cite{Fotchenkov} ($\bigtriangleup$).}
\end{figure}

The coefficients of piezoelectric strain $d_{1i}$ are shown in
figure~\ref{fig-d1i} at zero and high bias fields $E_1$.  The data
of \cite{Schmidt} were extracted from the given therein
values of the $d_{11}g_{11}$ product and of $g_{11}$. In absence
of external field $d_{1i}$ actually diverges at the Curie
temperatures, due to the term proportional to $s_{44}^E$ in
equation~(\ref{2.7}); the bias field smears out these anomalies and
lowers the peaks of the coefficients. A good agreement with
the experimental points is obtained.

\subsection{Hydrostatic pressure effects}

\begin{figure}[ht]
\centerline{\includegraphics[width=6.5cm]{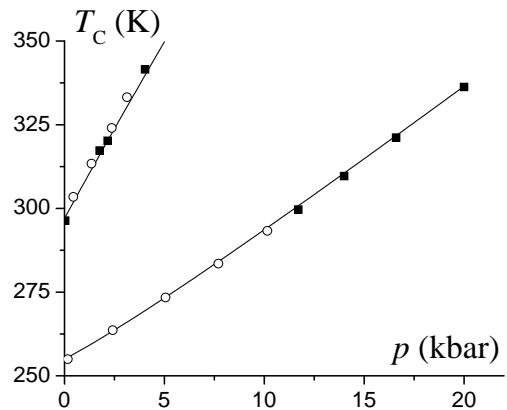}}
\caption{\label{fig-pressure-tc} Hydrostatic pressure dependence
of the Curie temperatures of Rochelle salt. Lines: a theory; open
and closed symbols: experimental points taken from \cite{Bancroft}
and \cite{Samara}.}
\end{figure}

Below we shall discuss how the high-pressure effects are described by the proposed modification of the Mitsui model.
Figure~\ref{fig-pressure-tc} shows the calculated hydrostatic
pressure dependence of the %
Curie
temperatures. The proposed theory
reproduces the experimentally observed linear increase of both
Curie temperatures with pressure at its low values, as well as the
increase of $\partial T_{\rm C1}/\partial p_{\mathrm{h}}$ at higher
pressures. The calculated slopes are 3.7~K/kbar at low pressures
and 4.3~K/kbar above 12~kbar.

To describe the pressure variation of the static permittivity we
need to make some changes in the theory parameters.  The above given
 value of $\mu_1$ provides a roughly equal fit to many
different experimental data for $\chi_{11}^\sigma$ in
paraelectric phases at atmospheric pressure. However, the sample
to sample variation of permittivity at ambient pressure reaches
10\% even in good samples \cite{35} and is much larger in samples
with defects. This is comparable with the changes in the Curie
constants produced by hydrostatic pressure \cite{Samara,ourrs}
below 10~kbar. Thus, it seems impossible to try to describe these
fine high-pressure effects, using for a non-deformed crystal the
averaged value of $\mu_1$ given in table~\ref{table-parameters}.
Instead, we shall determine separate values of $\mu_1$, which
provide the best fit to the experimental data for static
permittivity obtained
in \cite{Samara} and in \cite{ourexp} at each pressure considered therein.
Thus, the possible pressure variation of the effective dipole
moment $\mu_1$ can be inferred.

The measurements, reported both in \cite{Samara} and in
\cite{ourexp}, revealed a decrease of the peak value of
permittivity at a lower Curie temperature with increasing
pressure. This is attributed \cite{Samara} to a partial
clamping of the samples due to the increased viscosity of the
pressure-transmitting fluid and suppression of the piezoelectric
shear strain $\eps_4$. To calculate the dielectric permittivity of
a partially clamped crystal we assume that the clamping caused by
the viscous fluid under hydrostatic pressure is uniform throughout
the crystal sample. We assume that, at least, the part of shear
strain $\eps_4$ induced by the
measuring electric field (normally at 1 kHz) is smaller than the one given by
equation~(\ref{strain4}). The total strain is then equal to
\begin{equation}
\label{clampedstrain} \varepsilon _4=  \varepsilon _{s4} + k\left(
\frac{e_{14}^0}{c_{44}^{E0}} E_1 - \frac{2\psi _4
}{vc_{44}^{E0}}\xi_i\right).
\end{equation}
Here $\varepsilon _{s4}$ is a spontaneous part of the strain;
$\xi_i$ is the field-induced part of the order parameter.

The introduced phenomenological coefficient $0<k<1$ describes the
extent  to which the external pressure suppresses the shear strain
$\eps_4$: the cases $k=0$ and $k=1$ correspond to the totally
clamped and free crystals, respectively. Values of $k$ are,
naturally, pressure and temperature dependent and are different for
experimental setups with different pressure-transmitting liquids.
Thus, the level of clamping at $T_{\rm C1}$ was apparently much
higher in the experimental setup of \cite{ourexp} than in
\cite{Samara}, possibly because the corresponding temperatures are
lower, and the viscosity of pressure-transmitting liquid is
higher (benzine and silicone oil \cite{ourexp} vs pentane
isopentane mixture \cite{Samara}).

Substituting equation~(\ref{clampedstrain}) into equation~(\ref{2.4b}) we
obtain the corresponding polarization.
Differentiating it with respect to $E_1$, taking into account
equation~(\ref{ord-par}) and neglecting variation of diagonal strains
with the field, we get the dielectric susceptibility of a
partially uniformly clamped crystal
\begin{equation}\label{partiallyclamped}
    \chi_{11}^{k}=\chi_{11}^{k0}+\frac{\beta(\mu_1^k)^2}{2v\eps_0}
    \frac{\varphi_3}{\varphi_2-k\Lambda\varphi_3}\,,
\end{equation}
where
\[
\chi_{11}^{k0}=\chi_{11}^{\eps0}+\frac{k}{\eps_0}e_{14}^0d_{14}^0\,,
\qquad \mu_1^k=\mu_1-2k\psi_4d_{14}^0\,.
\]
From equation~(\ref{partiallyclamped}) in the limiting cases $k=0$ and
$k=1$ we obtain susceptibilities of totally clamped and free (in
the paraelectric phases) crystals. In the ferroelectric phase, a
more accurate expression for susceptibility should also
contain terms like $\sum{e_{1i}d_{1i}}$, albeit small,  produced
by contributions of diagonal strains. In this subsection these
contributions will be neglected, and permittivity will be
calculated using equation~(\ref{partiallyclamped}).

The temperature curve of dielectric permittivity near the
lower Curie point at atmospheric pressure presented in
\cite{Samara} appears to be drawn qualitatively. In the
fitting procedure we relied on the data of \cite{Lunk} for
these temperatures. Near the upper Curie point, the data of
\cite{Samara} and \cite{Lunk} agree fairly well.

\begin{wrapfigure}{i}{0.47\textwidth}
\centerline{\includegraphics[width=0.446\columnwidth]{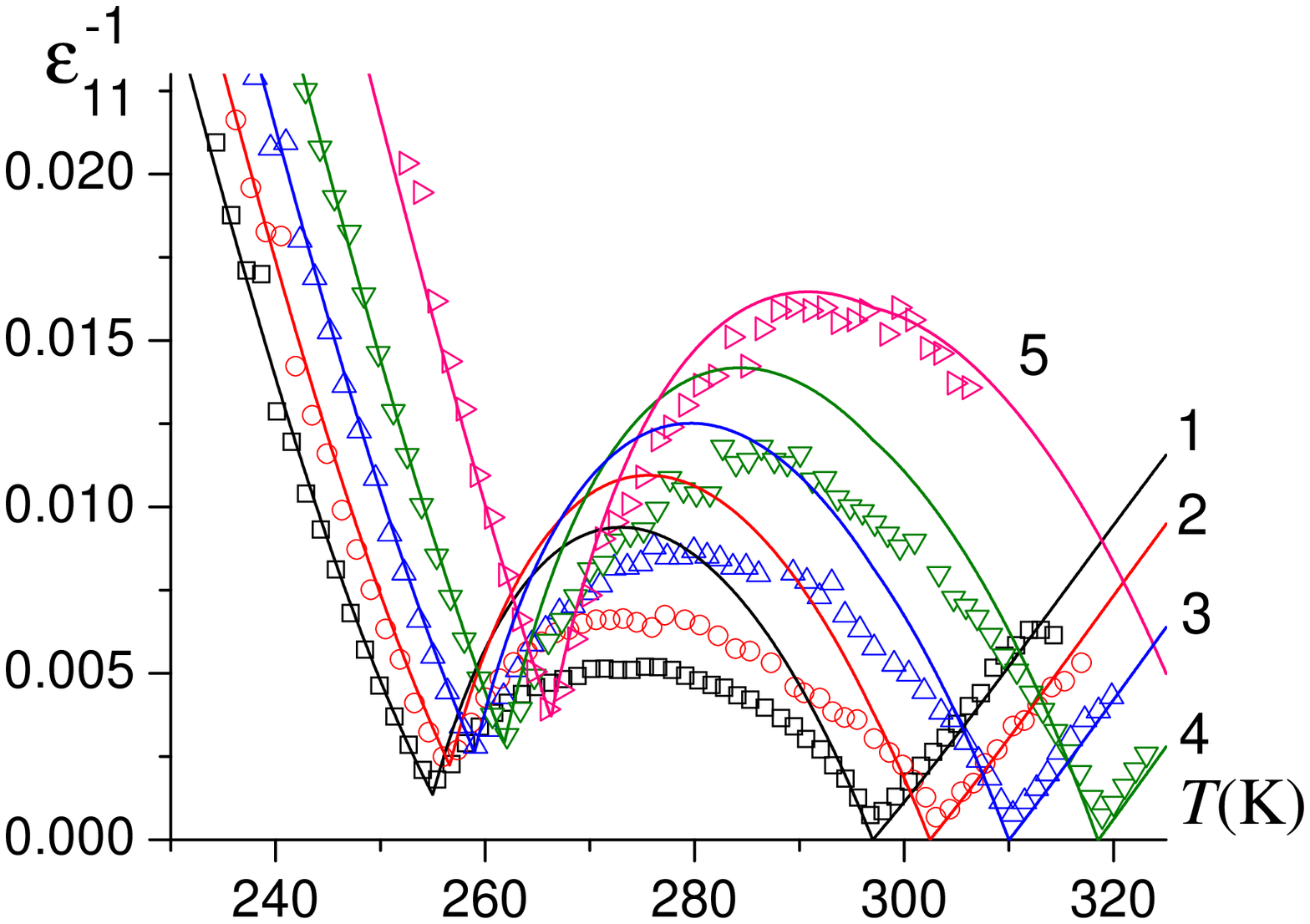}}%
\caption{\label{fig-pressure-slivka}  The temperature dependence
of the inverse static permittivity of Rochelle salt at different
values of hydrostatic pressure $p_{\mathrm{h}}$ (kbar): 1, $\square$: 0;  2,
$\bigcirc$: 0.5; 3, $\triangle$: 1.2; 4, $\nabla$: 2; 5,
$\vartriangleright$: 3.2. Lines: a theory; symbols: experimental
points taken from \cite{ourexp}.}
\end{wrapfigure}

\begin{figure}[ht]
\includegraphics[width=0.46\textwidth]{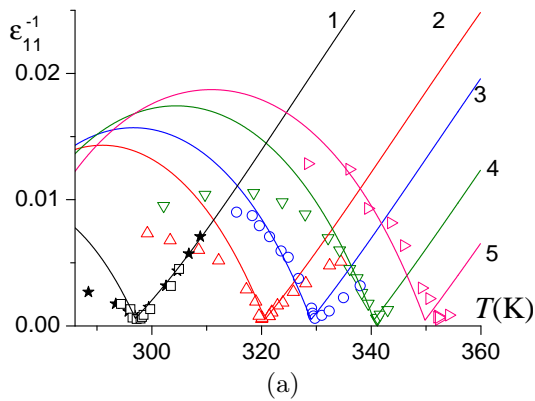}%
\hfill%
\includegraphics[width=0.46\textwidth]{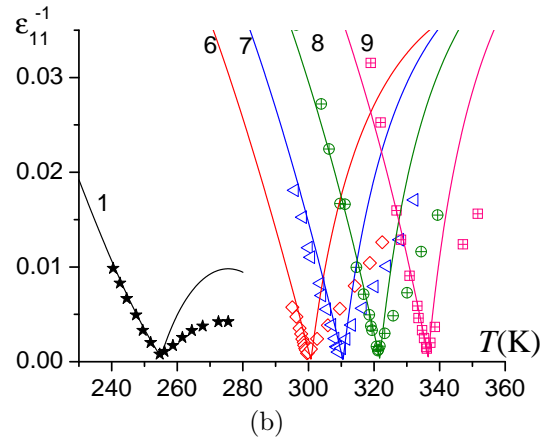}%
\\%
\parbox[t]{0.48\textwidth}{%
\centerline{(a)}%
}%
\hfill%
\parbox[t]{0.48\textwidth}{%
\centerline{(b)}%
}%
\caption{\label{fig-pressure-samara} (Color online) {The
temperature dependence of the inverse static permittivity of
Rochelle salt near the upper (left) and lower (right) Curie
points. The values of hydrostatic pressure are: 1, $\bigstar$,
$\square$: 0; 2, $\triangle$: 2.2; 3, $\bigcirc$: 3; 4,
$\triangledown$ : 4.1; 5, $\vartriangleright$: 5; 6, $\diamond$:
11.6; 7, $\vartriangleleft$: 14; 8, $\bigoplus$: 16.6; 9,
$\boxplus$: 20. Symbols are experimental points taken from
\cite{Lunk} ($\bigstar$) and \cite{Samara} (other symbols). Lines:
a theory.} }
\end{figure}

Comparison of the calculated temperature dependences of the
permittivity with experimental  data is given in
figures~\ref{fig-pressure-slivka} and \ref{fig-pressure-samara}. A
fairly good description of the experiment in paraelectric
phases is obtained at a proper choice of $k$ and $\mu_1$ values.
The disagreement at 2.2 and 4.1~kbar in
figure~\ref{fig-pressure-samara} is due to the mismatch between the
calculated and experimental Curie temperatures at these pressures.
The disagreement observed in the ferroelectric phase is due to the
essential domain contributions to the permittivity, not included
into the present model.

 To fit the data of \cite{ourexp} the coefficient $k$ is
taken to be 1 at 297~K (a free crystal). At 255~K we use the
following values of $k$: 0.65, 0.4, 0.25, 0.2 at 0, 0.5, 1.2,
2.0~kbar, respectively. A linear interpolation between these
values at 255~K and 1 at 297~K is used. For 3.2~kbar we use $k=0$
at 265 K and 1 at 297~K, also with a linear interpolation for the
intermediate temperatures. For the data of \cite{Samara} we
use $k= 0.9$ above 11.7~kbar for the lower Curie temperature.

The pressure and temperature variation of the dipole moment
$\mu_1$ is more complicated and contradictory. First, to fit the
experimental data  of \cite{ourexp} for permittivity at
ambient pressure we have to assume that $\mu_1$ \textit{increases}
with temperature; whereas for the data of \cite{Samara},
\cite{Lunk} it should be assumed to \textit{decrease} with increasing
temperature, but slower than it was given in
table~\ref{table-parameters}.

To fit the permittivity points at high pressures we need to assume
that the pressure dependence of $\mu_1$ is opposite to its
temperature dependence: if it decreases with increasing
temperature, then it increases with pressure and vice versa. Thus,
for the data of \cite{Samara} we use the following dependence
\[\mu_1=\mu_1^0\left[1+k_{\mathrm{T}}\left(T-T_{\rm C2}^0\right)\right]\left(1+k_{\mathrm{p}}p\right)\]
with $\mu_1^0=8.94\cdot10^{-30}$~C$\cdot$m and
$k_{\mathrm{T}}=-0.001$~K$^{-1}$. The pressure coefficient $k_{\mathrm{p}}$ was
$0.03$~kbar$^{-1}$ for pressures below 5~kbar and 0 above 10~kbar.

This behavior is quite unusual since the dipole moments are expected
to be reduced by hydrostatic pressure or by decreasing temperature
due to the overall reduction of interatomic distances. The
increase of $\mu_1$ with hydrostatic pressure can be explained
with the help of a certain assumption used in constructing the spatial
four-sublattice model of Rochelle salt \cite{Velychko}. It states
that the dipole moments in it are actually the 3D vectors, which
are not oriented along the $a$-axis like in the two-sublattice
model. Their projections on the $b$ and $c$ axes are different
from zero, but compensated at all temperatures, unless an electric
field perpendicular to spontaneous polarization is applied. The
$a$-projections are the dipole moments of a two-sublattice
Mitsui model, compensated in paraelectric phases. It may be
assumed that hydrostatic pressure rotates the spatial dipoles in
such a way that their projections on the $a$-axis (and $\mu_1$)
increase. This increase is fast at low pressures and slows down
above 10~kbar, possibly because the dipoles are already oriented
along the $a$-axis.

To fit the data of  \cite{ourexp} we use
$\mu_1^0=1.0\cdot10^{-30}$~C$\cdot$m, $k_{\mathrm{T}}=0.003$~K$^{-1}$, and
$k_{\mathrm{p}}=-0.012$~kbar$^{-1}$.  This is consistent with the picture
when pressure and temperature affect the dipole moment $\mu_1$ in
the opposite ways.  The  measurements \cite{ourexp} of humidity
(extreme drying and wetting) effect on the permittivity of
Rochelle salt have shown that in strongly wet samples the
permittivity increases significantly as compared to the permittivity of normal
samples, especially in the upper paraelectric phase. It appears
that the samples used in the hydrostatic pressure studies
\cite{ourexp} were moderately wet. On the other hand, a decrease of
$\mu_1$ with increasing temperature (like in \cite{Samara,35,Lunk}) seems to be an intrinsic behavior of normal
Rochelle salt samples. To explain the increase of $\mu_1$ with
temperature in wet samples we can assume that due to the excess of
water, the crystal conductivity increases, thus increasing the
dielectric permittivity, especially at high temperatures where the
mobility of charge carriers is very high.

\subsection{Uniaxial pressures effects}
To get a clearer picture of the uniaxial pressure effect on the
Curie temperature and dielectric permittivity, it is useful to
analyze it within the phenomenological approach. We start  with
the thermodynamic potential
\[
\frac{G_1}{V}=-\frac12\sum_{ij=1}^3s_{ij}p_ip_j-\frac{s_{44}^P}2p_4^2-P_1\sum_{i=1}^3M_{1i4}p_ip_4
+g_{14}P_1p_4+
P_1^2\sum_{i=1}^3Q_{1i}p_i+\frac12\alpha P_1^2 +\frac14\beta
P_1^4\,,
\]
(where
$s^P_{i4}=M_{1i4}P_1$ are the elastic compliances, and the
$M_{1i4}$ are simply the proportionality coefficients between
$s^P_{i4}$ and polarization. Also, $Q_{1i}$ are the
electrostriction constants; $\alpha=\alpha_{\mathrm{T}}(T-T_{\rm C2}^0)$ or
$\alpha=\alpha_{\mathrm{T}}(T_{\rm C1}^0-T)$, $\beta$ are the coefficients of
the Landau expansion). Hence, one gets the following
expressions for the temperature and magnitude of the permittivity
maxima
 \begin{eqnarray}
 &&
\label{phenom} T^{\mathrm{max}}_{1,2}=T_{\rm
C1,2}^0\pm\frac{2}{\alpha_{\mathrm{T}}}\sum_{i=1}^3Q_{1i}p_i
+\frac34\frac{(4\beta)^{1/3}}{\alpha_{\mathrm{T}}}
 \left(E_1-\sum_{ij=1}^3M_{1i4}p_ip_4+g_{14}p_4\right)^{2/3},\nonumber\\
 &&\varepsilon_{\mathrm{max}}^{-1}=\frac32(4\beta)^{1/3}
  \left(E_1-\sum_{ij=1}^3M_{1i4}p_ip_4+g_{14}p_4\right)^{2/3}.
\end{eqnarray}
In the absence of the fields conjugate to the order parameter ($E_1$
and $p_4$), the $T^{\mathrm{max}}_{1,2}$ correspond to the phase transition
temperatures. As one can see, in this approximation the uniaxial
and hydrostatic ($p_{\mathrm{h}}=p_1=p_2=p_3$) pressures lead to linear
shifts of the Curie temperatures, where the permittivity still
diverges. In combination with the shear pressure $p_4$ they smear
the transitions, shift the permittivity maxima, and lower the
peaks height.
 It also follows from equation~(\ref{phenom}) that if $p_4=0$
 \begin{equation}
 \label{eq1}
 \frac{\partial T_{{\rm C}k}}{\partial p_{\mathrm{h}}}=\sum_{j=1}^3\frac{\partial T_{{\rm
C}k}}{\partial p_j}\,. \end{equation}

In table~\ref{table-uniaxial-slopes} we summarize the experimental
data on the uniaxial pressure slopes of the transition
temperatures of Rochelle salt and  present our results obtained
within the herein developed modification of the Mitsui model. The
data of \cite{Unruh} are estimated from the presented therein
$\eps_{11}(T)$ curves, each being measured at a single value of $p_i$
near the upper Curie temperature.

\begin{table}[hbt]
  \caption{Uniaxial pressure derivatives of the transition temperatures of Rochelle salt (in K/kbar).}\label{table-uniaxial-slopes}
  \vspace{1ex}
\begin{center}
  \begin{tabular}{ccccc}
  \hline  \hline
 & \cite{Imai2} &
\cite{Unruh} & \cite{biaxial} & this work  \\
\hline
    $\partial T_{\rm C1}/\partial p_1$ & --29  &  & & --26.6  \\
    $\partial T_{\rm C1}/\partial p_2$ &  15  &  & &  14.8 \\
    $\partial T_{\rm C1}/\partial p_3$ &  17 &   & &   18.0 \\
    $\partial T_{\rm C1}/\partial(p_2+p_3)$ &  &   & $30\pm2$ &  34.0  \\
    $\partial T_{\rm C2}/\partial p_1$ &  35 &  35.6 && 32.9   \\
    $\partial T_{\rm C2}/\partial p_2$ &  --16 & --18  & & --16.5  \\
    $\partial T_{\rm C2}/\partial p_3$ &  --8 &  --6.7 & & --8.5  \\
    $\partial T_{\rm C2}/\partial(p_2+p_3)$ &  &   & $-22\pm1$ &  --26.2  \\
    \hline  \hline
  \end{tabular}
  \end{center}
\end{table}

The theory and experiment for the uniaxial pressures agree within
10\%, which is close to the experimental error \cite{Imai2}. The
theoretical dependence of the Curie temperatures on the biaxial
pressure $p_2+p_3$ is a little stronger than the experimental one.
Also it can be noticed that equation~(\ref{eq1}) is not fulfilled.

Independent measurements \cite{biaxial,ourexp,Unruh} of
dielectric susceptibility of Rochelle salt under different
uniaxial and biaxial pressures revealed a decrease of the peak
values of susceptibility at the Curie points as well as smearing out
of the peaks. These effects are enhanced with increasing pressures.

A uniform partial clamping of samples by an apparatus creating the
uniaxial pressures can explain the observed lowering of the peaks,
but not the smearing of the transitions. As an intrinsic
phenomenon, both these effects can be caused by application of an
external field conjugate to the order parameter: the electric
field $E_1$ directed along the axis of spontaneous polarization or
the shear stress $\sigma_4$, that is, by the field which induces
polarization $P_1$ in the paraelectric phases. In an ideal experiment, no uniaxial or
biaxial pressure applied along the orthorhombic crystallographic
axes should act in this way, because piezoelectric coefficients associated with these pressures are
zeros outside the ferroelectric phase.

It appears that the observed smearing of the anomalies is an
artefact caused by experimental errors. We can think of the
following factors that in a real experiment can lead to the
smearing.

\begin{itemize}
\item[(i)]  Stress inhomogeneity. Even a weak inhomogeneity of the applied
pressure, partial clamping/sample end constraints, surface
irregularity result in a non-uniform strain distribution over the
crystal sample; thus, in different parts of the sample the phase
transition is shifted to different temperatures. The higher is
pressure, the larger is the difference between these temperatures,
and the more diffuse is the transition, exactly as observed in
\cite{ourexp}.
\item[(ii)]  Stray shear stress $\sigma_4$, whose effect would be
enhanced by its combination with the uniaxial pressures $p_i$, as
it follows from equation~(\ref{phenom}). The stress $\sigma_4$ can arise
even at a slight misorientation of the samples, as a component of
the uniaxial loading intended to create $p_2$ or $p_3$ pressures.
There can also occur built-in local shear stresses $\sigma_4$ caused
by sample defects, e.g. dislocations.
\end{itemize}

There will be too much uncertainty if we try to take the effect
of these factors into account in the theory. Thus, we shall not
attempt to describe the behavior of the static dielectric
susceptibility in uniaxially stressed Rochelle salt crystals.

\section{Concluding remarks}
We proposed a generalization of the deformable Mitsui model
\cite{ourrs}, which along with the piezoelectric shear strain
$\varepsilon_4$ takes into account the diagonal strains
$\eps_1$, $\eps_2$, and $\eps_3$ as well. In contrast to the previous
attempt \cite{monoclinic}, in order to incorporate the diagonal strains into
this model, the thermal expansion is consistently taken into account.

In the mean field approximation we find polarization and the
strains, as well as thermal,
elastic, and piezoelectric characteristics related to diagonal strains.
For the case of Rochelle salt we suggest an elaborated fitting procedure and
choose the set of the model parameter values, providing as good as possible
consistent description of all these characteristics, as well as of external
hydrostatic and uniaxial pressure effects.

The expression for susceptibility of a partially clamped
crystal is derived in order to describe the behavior of the observed
susceptibility of Rochelle salt under hydrostatic pressure near
the lower Curie point. By fitting the theoretical curves to the
experimental points, the pressure variation of the effective dipole
moment $\mu_1$ is estimated. An increase of $\mu_1$ with hydrostatic
pressure at low pressures and a decrease with increasing temperature
seem to be an intrinsic behavior for this crystal. The
interaction constants and the asymmetry parameter were found to
increase with hydrostatic pressure too. This is consistent with
the picture, where the dipole moments in Rochelle salt are 3D
vectors \cite{Velychko}, assuming they rotate under pressure in
such a way that their projection on the $a$-axis increases.

Apart from Rochelle salt, the developed modification of the model
with appropriate changes can be used for consideration of the
pressure effects and thermal expansion in other ferroelectric
crystals described by the Mitsui model. The fitting procedure,
however, for each such crystal will require extensive experimental
data; further measurements will, therefore, be necessary.

The presented model and the found values of its parameters are a
good starting point for developing a model description of
ferroelectricity in nanosize inclusions of compounds, to which the
Mitsui model is applicable, grown in a porous matrix
\cite{YadlovkerBerger,Baryshnikov1,Baryshnikov2}.

\ukrainianpart
\title{Модель Міцуї з діагональними деформаціями: об'єднаний опис впливу зовнішніх тисків
і теплового розширення в сеґнетовій солі  NaKC$_4$H$_4$O$_6\cdot4$H$_2$O}
\author{А.П. Моїна\refaddr{label1}, Р.Р. Левицький\refaddr{label1}, І.Р. Зачек\refaddr{label2}}

\addresses{
\addr{label1}Інститут фізики конденсованих систем НАН України,
79011 Львів, вул. Свєнціцького, 1
 \addr{label2} Національний університет ``Львівська політехніка'', 79013 Львів, вул. С.~Бандери, 12}
\makeukrtitle

\begin{abstract}
Запропоновано модифікацію деформівної двопідґраткової моделі Міцуї робіт
[Levitskii~R.R. et al, Phys. Rev.~B. 2003, \textbf{67}, 174112] та [Levitskii~R.R. et al.,  Condens. Matter
Phys., 2005, \textbf{8}, 881], яка послідовно враховує діагональні
компоненти тензора деформацій, що виникають під дією зовнішніх тисків
чи внаслідок теплового розширення. Розраховано пов'язані з цими деформаціями теплові, п'єзоелектричні та пружні характеристики
системи. Використовуючи запропоновану схему, для кристалів  сеґнетової солі знайдено такий набір параметрів теорії, що забезпечує
задовільне узгодження з експериментальними даними для залежностей температур Кюрі від гідростатичного та одновісних тисків, а також температурних
залежностей теплових деформацій, лінійних коефіцієнтів теплового розширення, пружних сталих $c_{ij}^E $ і
$c_{i4}^E $, п'єзоелектричних коефіцієнтів $d_{1i}$ і $g_{1i}$
($i=1,2,3$). Залежності діелектричної проникності від гідростатичного тиску описано за допомогою отриманого в роботі
виразу для проникності частково затиснутого кристалу. Виявлено, що дипольні моменти та параметр асиметрії в сеґнетовій солі зростають з гідростатичним тиском.

\keywords сеґнетова сіль, теплове розширення, гідростатичний тиск, одновісний тиск, модель Міцуї

\end{abstract}

\begin{thebibliography}{99}

\bibitem{80}
  Mitsui~T.,   Phys. Rev., 1958, {\bf  111}, 1259; \bibdoi{10.1103/PhysRev.111.1259}.

\bibitem{vaks-review}
  Vaks~V.G.,   Zinenko~V.I.,   Schneider~V.E.,  Sov. Phys. Uspekhi, 1983, \textbf{26}, 1059; \\\bibdoi{10.1070/PU1983v026n12ABEH004584}.

\bibitem{Lunk}
 Schneider~U.,   Lunkenheimer~P.,   Hemberger~J.,  Loidl~A.,
Ferroelectrics, 2000, \textbf{242}, 71; \\\bibdoi{10.1080/00150190008228404}.

\bibitem{Alexandrov}
 Aleksandrov~K.S.,  Anistratov~A.T., Ferroelectrics, 1976, \textbf{12}, 191; \bibdoi{10.1080/00150197608241423}.

\bibitem{Blat}  Blat~D.Kh.,  Zinenko~V.I.,   Fiz. Tverd. Tela (Leningrad), 1976, \textbf{18}, 3599.


\bibitem{watarai1}
 Watarai~S.,    Matsubara~T.,  J. Phys. Soc. Jpn., 1978, \textbf{45}, 1807; \bibdoi{10.1143/JPSJ.45.1807}.

\bibitem{watarai2}
 Watarai~S.,    Matsubara~T.,  Sol.  State Comm., 1980, \textbf{35}, 619; \bibdoi{10.1016/0038-1098(80)90595-5}.

\bibitem{sasd}
Korynevskii~N.A.,  Ferroelectrics, 2002, \textbf{268}, 207; \bibdoi{10.1080/00150190211058}.

\bibitem{ourrs}
 Levitskii~R.R.,   Zachek~I.R.,   Verkholyak~T.M.,   Moina~A.P.,  Phys. Rev.~B, 2003, \textbf{67}, 174112; \\ \bibdoi{10.1103/PhysRevB.67.174112}.

\bibitem{ourrs2}
 Moina~A.P.,   Levitskii~R.R.,   Zachek~I.R.,  Phys. Rev.~B, 2005,
\textbf{71}, 134108; \\\bibdoi{10.1103/PhysRevB.71.134108}.

\bibitem{YadlovkerBerger}
  Yadlovker~D.,   Berger~S.,  Phys. Rev. B, 2005,  \textbf{71}, 184112; \bibdoi{10.1103/PhysRevB.71.184112}.

\bibitem{YadlovkerBerger2}
Yadlovker~D.,   Berger~S.,  J. Electroceram., 2007, \textbf{22}, 281.

 \bibitem{Baryshnikov1}
 Baryshnikov~S.V.,   Charnaya~E.V.,   Stukova~E.V.,  Milinskii~A.Yu.,
 Cheng Tien, Fiz. Tverd. Tela, 2010, \textbf{52}, 1347 [Phys.
Solid State  \textbf{52},  1444 (2010); \bibdoi{10.1134/S1063783410070206}].

 \bibitem{Baryshnikov2}
Cheng Tien,  Charnaya~E.V., et al.,  J. Phys.: Condens. Matter, 2008,
\textbf{20}, No.~21, 215205; \bibdoi{10.1088/0953-8984/20/21/215205}.


\bibitem{rbhso4nano}
 Niznansky~D.,   Plocek~J.,   Svobodova~M.,   Nemec~I.,  Rehspringer~J.-L.,  Vanek~P.,   Micka~Z., J. Sol-Gel Sci. Technol., 2003, \textbf{26},  447; \bibdoi{10.1023/A:1020702005541}.

\bibitem{Morozovska}
  Morozovska~A.N.,   Eliseev~E.A.,   Glinchuk~M.D.,  Phys. Rev. B, 2006, \textbf{73}, 214106; \\\bibdoi{10.1103/PhysRevB.73.214106}.

\bibitem{AAY}
 Levitskii~R.R., Moina~A.P.,   Andrusyk~A.Ya.,   Slivka~A.G.,   Kedyulich~V.M.,  J. Phys. Stud., 2008, \textbf{12}, 2603.

\bibitem{SASD-pressure}
 Lipinski~I.E.,  Kuriata~J.,  Korynevskii~N.A.,  Ferroelectrics, 2005, \textbf{317}, 115.

\bibitem{monoclinic}
 Levitskii~R.R.,   Zachek~I.R.,   Moina~A.P.,  Condens. Matter Phys., 2005,
\textbf{8}, 881.

\bibitem{Suzuki}
 Suzuki~E.,   Shiozaki~Y.,  Phys. Rev. B, 1996,  {\bf 53}, 5217; \bibdoi{10.1103/PhysRevB.53.5217}.

\bibitem{Petzelt3}
Hlinka~J.,   Kulda~J.,    Kamba~S.,  Petzelt~J., Phys. Rev. B, 2001,  {\bf 63}, 052102; \\\bibdoi{10.1103/PhysRevB.63.052102}.

\bibitem{Bancroft}
Bancroft~D., Phys. Rev., 1938, \textbf{53}, 587; \bibdoi{10.1103/PhysRev.53.587}.

\bibitem{Samara}
 Samara~G.A.,  J. Phys. Chem. Solids, 1965,  \textbf{26}, 121; \bibdoi{10.1016/0022-3697(65)90079-X}.

\bibitem{Imai2}
Imai~K.,  J. Phys. Soc. Jpn., 1975, \textbf{39}, 868; \bibdoi{10.1143/JPSJ.39.868}.

\bibitem{Unruh}
Unruh~H.-G.,   M\"{u}ser~H.E.,  Ann. Phys., 1967, \textbf{474}, 28; \bibdoi{10.1002/andp.19674740105}.

\bibitem{biaxial}
Mori~K.,   Hayashi~M., J. Phys. Soc. Jpn., 1972, \textbf{33}, 1396; \bibdoi{10.1143/JPSJ.33.1396}.

\bibitem{Schmidt}
Schmidt~G.,  Z. Angew. Phys., 1961, \textbf{161},  579; \bibdoi{10.1007/BF01341554}.

\bibitem{Fotchenkov}
Fotchenkov~A.A.,  Sov. Phys. Crystallogr., 1960, \textbf{5}, 390.

\bibitem{Sailer}
Sailer~E.,  Unruh~H.-G., Ferroelectrics, 1976,  \textbf{12},  285; \bibdoi{10.1080/00150197608241452}.

\bibitem{Bronowska}
 Wlodarz~M.,   Bronowska~W.,  Dziedzic~J.,  Ferroelectr. Lett., 1988,
\textbf{9}, 83; \\\bibdoi{10.1080/07315178808200706}.

\bibitem{Imai1}
Imai~K.,  J. Phys. Soc. Jpn., 1976, \textbf{41}, 2005; \bibdoi{10.1143/JPSJ.41.2005}.

\bibitem{Landolt}
Numerical Data and Functional Relationships in Science and
Technology, eds. Hellwege~K.-H. and Hellwege~A.M.
Landolt-Bornstein, New Series. Group III: Crystal and Solid State
Physics, Vol. 16, Pt.~b. Springer-Verlag, Berlin, 1982.

\bibitem{AHV}
Anderson~P.W., Halperin~B.I., Varma~C.M.,   Philos. Mag., 1972, \textbf{25}, 1; \bibdoi{10.1080/14786437208229210}.

\bibitem{Phillips}
Phillips~W.A., J. Low Temp. Phys., 1972, \textbf{7}, 351; \bibdoi{10.1007/BF00660072}.

\bibitem{9}
Bronowska~W.J., J. Appl. Crystallogr., 1981, {\bf 14}, 203; \bibdoi{10.1107/S0021889881009114}.

\bibitem{vaks}
Vaks~V.G.,  Introduction into Microscopic Theory of
Ferroelectrics. Nauka, Moscow, 1973 (in Russian).

\bibitem{werch}
Levitskii~R.R., Verkholyak~T.M., Kutny~I.V.,   Hil~I.G. Preprint arXiv:cond-mat/0106351~(unpublished).

\bibitem{Dublenych}
Dublenych~Yu.I.,  Condens. Matter Phys., 2011,  \textbf{14}, 23603; \bibdoi{10.5488/CMP.14.23603}.

\bibitem{64B2}
Berlincourt~D.A.,   Curran~D.R.,  Jaffe~H. -- In: Physical Acoustics,
Vol.~1, Part A, p.~169--270, \linebreak ed. W.P.~Mason. Academic
Press, New York, 1964.


\bibitem{dosSantos}
dos Santos~A.,   de Menezes~A.S., Sasaki~J.M., Cardoso~L.P., Acta Crystallogr., Acta Crystallogr., Sect. A: Found. Crystallogr., 2008, \textbf{64}, C544.

\bibitem{ourexp}
Slivka~A.G.,   Kedyulich~V.M.,   Levitskii~R.R.,   Moina~A.P.,   Romanyuk~M.O.,
Guivan~A.M.,  Condens. Matter Phys., 2005, \textbf{8}, 623.

\bibitem{35}
Sandy~F., Jones~R.V., Phys. Rev., 1968, {\bf 168}, 481; \bibdoi{10.1103/PhysRev.168.481}.

\bibitem{Tatsumi}
Tatsumi~M.,    Matsuo~T.,    Suga~H.,    Seki~S.,  J. Phys. Chem. Solids, 1978,
\textbf{39}, 427; \bibdoi{10.1016/0022-3697(78)90084-7}.

\bibitem{Velychko}
Stasyuk~I.V.,   Velychko~O.V.,  Ferroelectrics, 2005, \textbf{316}, 51; \bibdoi{10.1080/00150190590963138}.

\end{thebibliography}
\end{document}